\documentclass[aps,prd,reprint,superscriptaddress,nofootinbib,longbibliography]{revtex4-2}
\usepackage{amssymb,amsmath,bm,natbib,braket}
\usepackage[dvipsnames]{xcolor}
\usepackage[colorlinks = true,
            linkcolor = Maroon,
            urlcolor  = Maroon,
            citecolor = Maroon]{hyperref}
\usepackage{microtype}
\usepackage{graphicx}
\usepackage{xspace}
\usepackage{bm}
\usepackage{cancel}
\usepackage{bbm}
\usepackage{tikz}
\usepackage{dsfont}

\usetikzlibrary{decorations.markings}
\usetikzlibrary{decorations.pathmorphing}
\tikzset{snake it/.style={decorate, decoration=snake}}
\usetikzlibrary{matrix,calc}
\definecolor{harvardcrimson}{rgb}{0.79, 0.0, 0.09}

\begin{document}

\title{An EFT for anisotropic antiferromagnets: gapped Goldstones, pseudo-Goldstones, and phase transitions}

\author{Pier~Giuseppe~Catinari}
\email{piergiuseppe.catinari@uniroma1.it}
\affiliation{Dipartimento di Fisica, Sapienza Universit\`a di Roma, Piazzale Aldo Moro 2, I-00185 Rome, Italy}
\affiliation{INFN Sezione di Roma, Piazzale Aldo Moro 2, I-00185 Rome, Italy}

\author{Angelo~Esposito}
\email{angelo.esposito@uniroma1.it}
\affiliation{Dipartimento di Fisica, Sapienza Universit\`a di Roma, Piazzale Aldo Moro 2, I-00185 Rome, Italy}
\affiliation{INFN Sezione di Roma, Piazzale Aldo Moro 2, I-00185 Rome, Italy}

\author{Shashin~Pavaskar}
\email{pavaskar@illinois.edu}
\affiliation{Department of Physics, University of Illinois at Urbana-Champaign, Urbana, Illinois 61801, USA}
\affiliation{IQUIST, University of Illinois at Urbana-Champaign, Urbana, Illinois 61801, USA}

\date{\today}

\begin{abstract}
\noindent We build and discuss a low energy effective field theory for anisotropic antiferromagnets in the presence of an external magnetic field. Such an effective theory is simple yet rich and features a number of phenomena such as the appearance of gapped Goldstones, pseudo-Goldstones, and a first order ``spin-flop'' phase transition, all within the regime of validity of the theory. We also discuss in detail, the quantization procedure of the free theory in the presence of a magnetic field, which is made non-trivial by the presence of a single-time derivative term. 
This class of materials makes a precious test field for exotic phenomena in quantum field theory. Moreover, we explicitly perform the matching of the effective theory to the short distance theory of a specific antiferromagnet, namely, nickel oxide. The latter is particularly relevant in light of recent proposals of employing this material towards the hunt for light dark matter. As a byproduct of our study, we also re-evaluate the role played by discrete symmetries in magnetic materials, presenting it in a way that is completely consistent with the proper low energy EFT ideology.
\end{abstract}

\maketitle

%%%%%%%%%%%%%%%%%%%%%%%%%%%%%%%%%%%%%%%%%%%%
%%%%%%%%%%%%%%%%%%%%%%%%%%%%%%%%%%%%%%%%%%%%

\section{Introduction}

\noindent Antiferromagnets are magnetic materials which, below a certain temperature, exhibit long range order but no net magnetization. Contrary to ferromagnets, they were initially thought to have limited practical applications~\cite{neel1971magnetism}. Recent years, however, have witnessed the development of a plethora of such applications, ranging from electron transport controlled by the spin degrees of freedom~\cite[e.g.,][]{baibich1988giant,binasch1989enhanced}, the so-called spintronics, the development of quantum sensors~\cite[e.g.,][]{lachance2019hybrid,doi:10.1126/sciadv.1603150}, and, recently, light dark matter detection~\cite{EspositoPavaskar,Marsh:2018dlj}.
Moreover, magnetic materials, both ferromagnets and antiferromagnets, are typically regarded as the textbook example for the phenomenon of spontaneous symmetry breaking and the emergence of Goldstone bosons~\cite[e.g.,][]{Burgess:1998ku}. These are indeed nothing but the spin collective excitations of these systems, the so-called magnons.

\vspace{1em}

Yet, in reality, the spectrum of most antiferromagnets feature tiny but non-zero gaps. This can be attributed to magneto-crystalline anisotropies, which provide small effects of explicit symmetry breaking.
This is due to the presence of certain energetically favorable directions for the spins of the material, which are dictated by the underlying crystal structure of the antiferromagnet. These anisotropies, along with the exchange interaction, in turn determine the ground state of the system as well its behavior under the application of an external magnetic field.

\vspace{1em}

In this work we extend the well known effective field theory (EFT) for magnons in antiferromagnet~\cite[e.g.,][]{Burgess:1998ku,chandra1990quantum,Pavaskar_2022} to include these anisotropic effects.
In particular, we will consider the case where the system features two distinct anisotropies, the so-called ``easy plane'' antiferromagnets~\cite{rezende_introduction_2019}. The other class of anisotropic antiferromagnets, so-called ``easy axis'', feature only one anisotropy, but their spectrum can be obtained from that the previous class, simply setting one of the parameters to zero. We will show that this extended EFT, while being eventually rather simple, features a number of interesting phenomena, such as the presence of the so-called gapped Goldstones~\cite[e.g.,][]{Nicolis:2012vf,Nicolis:2013sga,Watanabe:2013uya,Cuomo:2020gyl}, their lifting to pseudo-Goldstones, as well as the presence of the so-called ``spin flop'' phase transition triggered by a varying external magnetic field, where the spin alignment in the ground state suddenly changes its direction~\cite[e.g.,][]{rezende_introduction_2019}. All this, is computable within the regime of validity of the EFT.

Moreover, in the presence of a magnetic field, as we will see, the theory cannot be diagonalized by a local redefinition. Hence it requires a careful quantization procedure, which we address in this work. Finally, we match our EFT to the short distance theory for a particular antiferromagnet, namely, nickel oxide (NiO). The latter, thanks to its simple structure and spin interactions, is often considered as the prototypical room-temperature antiferromagnet. Indeed, this material has been used to study a number of different processes, such as inelastic light scattering~\cite[e.g.,][]{grimsditch1998unexpected,milano2004effect} and magnetic response to terahertz frequencies~\cite{sievers1963far,satoh2010spin}. Moreover, as mentioned, it has recently been realized that such a system offers an optimal target to look for the scattering of dark matter with spin-dependent interactions and masses as low as the keV~\cite{EspositoPavaskar}, as well as to look for axion-like dark matter with masses in the meV range~\cite{pheno}.

\vspace{1em}
The goal of this work is also to present everything in a framework which can bridge the linguistic gap that is often present between the standard condensed matter and high energy physics literature. This is particularly important in light of the role that antiferromagnets might play in future searches for light dark matter, a genuinely high energy physics question which, however, requires condensed matter tools to be tackled.
Moreover, an EFT formulation has the great advantage of allowing to neatly disentangle those phenomena that are universal, i.e. solely due to long-distance physics, from those that are, instead, due to a specific short-distance realization of the system.

\vspace{1em}

\noindent{\it Conventions:} We adopt natural units, $\hbar = c = 1$, and an index notation such that $i,j,k = 1,2,3$ label spatial coordinates. We also employ Einstein's notation for summing over repeated indices.

%%%%%%%%%%%%%%%%%%%%%%%%%%%%%%%%%%%%%%
%%%%%%%%%%%%%%%%%%%%%%%%%%%%%%%%%%%%%%

\section{The EFT} \label{sec:EFT}

We now proceed to the construction of our EFT step-by-step, from the simplest instance to the most involved one. Throughout our manuscript, we assume to work at zero temperature. This is partially motivated by the phenomenological reasons highlighted in Ref.~\cite{pheno}, but also due to the fact that the inclusion of thermal effects in a general quantum field theory is typically rather non-trivial and beyond the scope of this work.

\subsection{The simplest EFT: the isotropic case} \label{sec:isotropic}

\noindent Let us start by reviewing the construction of the standard EFT for antiferromagnets, which has been discussed, for example, in Refs.~\cite{chandra1990quantum,Burgess:1998ku,Pavaskar_2022}. We will go through this review rather thoroughly, in order to set the stage to the subsequent study, as well as highlight a few points which can make an easier connection between the high and low energy formalisms used to described antiferromagnets.

At distances comparable to its lattice spacing, an antiferromagnet is essentially a collection of magnetic moments, which antialign with respect to the neighboring ones. These magnetic moments typically arise from the fundamental spin of the unpaired electrons localized around each atom of the crystal. The exchange and superexchange interactions between electrons pertaining to different atoms give rise to a coupling between these spins. The simplest example of a short distance Hamiltonian describing such a system is given by the Heisenberg model~\cite[e.g.,][]{lovesey},
\begin{align} \label{eq:H0}
    H_0 = \sum_{\bm i, \bm j \in \mathds{A}} J_{\bm i\bm j} \, \bm{\mathcal{S}}_{\bm{i}} \cdot \bm{\mathcal{S}}_{\bm j} + \sum_{\bm i, \bm j \in \mathds{B}} J_{\bm i\bm j} \, \bm{\mathcal{S}}_{\bm{i}} \cdot \bm{\mathcal{S}}_{\bm j} \,,
\end{align}
where $\bm i$ and $\bm j$ are positions on the lattice, which is usually seen as ``bipartite'', i.e., as composed of two sublattices, which we label as $\mathds{A}$ and $\mathds{B}$. $\bm{\mathcal{S}}_{\bm i}$ is the spin pertaining to a given lattice site, and $J_{\bm i \bm j}$ are the above mentioned couplings. 
For an antiferromagnet, all the spins have the same magnitude, $|\bm{\mathcal{S}}_{\bm i}| = \mathcal{S}$. 

In an antiferromagnet, the couplings $J$ are such that the zero-temperature classical ground state, i.e.,~the state corresponding to the minimum classical energy,\footnote{Contrary to ferromagnets, the exact {\it quantum} ground state of an antiferromagnet is unknown, and it is only approximated by its classical counterpart. Yet, the spin wave theory built out of this latter ground state gives a very accurate description of experimental results~\cite[e.g.,][]{lovesey}.} corresponds to a configuration where all the spins in the sublattice $\mathds{A}$ are aligned parallel to each other, and those in the sublattice $\mathds{B}$ are aligned in the opposite direction. The emergence of this ground state is quantified by an order parameter, the N\'eel vector, given by $\bm{\mathcal{N}} \equiv \sum_{\bm i \in \mathds{A}} \bm{\mathcal{S}}_{\bm i} - \sum_{\bm i \in \mathds{B}} \bm{\mathcal{S}}_{\bm i}$. On the classical ground state, the N\'eel vector acquires an expectation value parallel to the spins, $\langle \bm{\mathcal{N}} \rangle \neq 0$, which can in principle be aligned in {\it any} direction. The spin-wave excitations of the system, called magnons, are the long wavelength modulations of the order parameter around its equilibrium value.

From a symmetry viewpoint, the Hamiltonian~\eqref{eq:H0} enjoys a global SO(3) symmetry, corresponding to the simultaneous rotation of all spins, $\bm{\mathcal{S}}_{\bm i} \to \mathcal{R} \, \cdot \, \bm{\mathcal{S}}_{\bm i}$, with $\mathcal{R}$ a $3\times3$ rotation matrix. Nonetheless, this symmetry is not respected by the ground state order parameter, which selects a specific direction. In other words, the original spin rotations are spontaneously broken down to the subgroup that leaves the order parameter unchanged, i.e., ${\rm SO}(3) \to {\rm SO}(2)$. In this respect, magnons are nothing but the associated Goldstone bosons and, as such, their dynamics can be described by an EFT valid at low energies and long wavelengths. 

Beside the continuous symmetry just discussed, a central role is played by two discrete symmetries: time reversal, $\mathcal{T}$, and a discrete rotation of $180^\circ$ in the plane where the spins lie, which we represent by $\mathcal{R}_\pi$. Both these transformations act on the spins by changing the direction of all of them. The order parameter thus changes sign under the separate action of $\mathcal{T}$ and $\mathcal{R}_\pi$, i.e.,
\begin{align}
    \bm{\mathcal{N}} \xrightarrow[]{\mathcal{T}} - \bm{\mathcal{N}} \,, \qquad \bm{\mathcal{N}} \xrightarrow[]{\mathcal{R}_\pi} - \bm{\mathcal{N}} \,.
\end{align}
Consequently, the theory must be invariant under their combined action, which leaves the order parameter unchanged.\footnote{In the literature, instead of the discrete rotation we are using here, one often finds a discussion involving the action of a translation by a single lattice site~\cite[e.g.,][]{Burgess:1998ku}, which effectively exchanges the two sublattices, $\mathds{A}$ and $\mathds{B}$, thus changing the sign of the order parameter}. Here we prefer to phrase everything in terms of the rotation for two reasons: (1) it is better defined at long wavelengths, where one is insensitive to distances comparable to the lattice spacing, and (2) we believe that, phrased this way, it allows for a unified description of ferromagnets and antiferromagnets, as we explain in Appendix~\ref{app:discrete}. 
The complete symmetry breaking pattern is then ${\rm SO}(3) \times \mathcal{T} \to {\rm SO}(2) \times (\mathcal{T} \, \mathcal{R}_\pi)$. 
The EFT we present here is built in order to implement this pattern.

\vspace{1em}

Now, magnons correspond to local modulations of the order parameter away from equilibrium, represented by a field $\hat{\bm n}(x)$, with $\langle \hat{\bm n} \rangle \propto \langle \bm{\mathcal{N}} \rangle$ and $\hat{\bm n}^2 = 1$.\footnote{This can be understood from the fact that the coset space ${\rm SO}(3)/{\rm SO}(2) \simeq {\rm S}^2$, which is the unit 2-sphere, and $\hat{\bm n}$ denotes the unit vector on ${\rm S}^2$.} 
With this at hand, the most general low energy EFT for magnons with at most two derivatives is simply given by~\cite{Burgess:1998ku},
\begin{align}\label{eq:zero Lagrangian}
    \begin{split}
        \mathcal{L}_0={}&  \frac{c_1}{2}\left[{(\partial_t\hat{\bm{n}})}^2-v_\theta^2{(\nabla_i\hat{\bm{n}})}^2\right] \,,
    \end{split}
\end{align}
where $c_1$ is an effective coefficient that can be determined in terms of parameters of the microscopic Hamiltonian, as done in Ref.~\cite{EspositoPavaskar}, while $v_\theta$ is the magnon propagation speed.
All possible additional terms necessarily contain more derivatives and are therefore suppressed in the low energy/long wavelength limit.

Note that, by treating time derivatives and spatial gradients as completely independent from each other, we assumed the {\it explicit} breaking of boost invariance (whether Galileo or Lorentz is irrelevant). This is ultimately not what happens in Nature, as boosts are always {\it spontaneously} broken by the underlying lattice, and the ordinary gapless acoustic phonons are the corresponding Goldstones bosons necessary to recover the full symmetry, which they realize non-linearly~\cite[e.g.,][]{Leutwyler:1996er,Nicolis:2015sra}. Our approximation is valid as long as we are not interested in the phonon dynamics and their interplay with magnons. For an EFT including both, see Ref.~\cite{Pavaskar_2022}. 

\vspace{1em}

In the presence of an external uniform magnetic field, $\text{\bf H}$, the microscopic Heisenberg Hamiltonian in Eq.~\eqref{eq:H0} is modified by the Zeeman term, 
\begin{align}
    H_{\rm H} = H_0 - \mu \sum_{\bm i} \text{\textbf H} \cdot \bm{\mathcal{S}}_{\bm i} \,,
\end{align}
where $\mu$ is the gyromagnetic ratio of the spins. At the level of the EFT, as explained for example in Ref.~\cite{Brauner:2024juy}, this can be included by promoting the time derivative in the Lagrangian $\mathcal{L}_0$ to a ``covariant'' derivative, $\partial_t \hat{\bm n} \to \partial_t \hat{\bm n} + \mu \text{\bf H} \times \hat{\bm n}$, i.e.,
\begin{align}\label{eq:Lagrangian for B field}
    \mathcal{L}_{\rm H}=\frac{c_1}{2}\left[{(\partial_t\hat{\bm{n}}+\mu\text{\bf H}\times\hat{\bm{n}})}^2\,-{ v_\theta^2}{(\nabla_i\hat{\bm{n}})}^2\right]\,.
\end{align}
The correctness of this procedure is guaranteed by the fact, when going to the associated Hamiltonian density, one indeed recovers a purely Zeeman term, as expected.

To determine the spectrum of this theory, we first need to find the background value for $\hat{\bm n}$. This is done by minimizing the Hamiltonian density, taken in the static and homogeneous limit, which reads
\begin{align} \label{eq:scHB}
    \begin{split}
        \mathcal{H}_{\rm H}\big|_{\rm{stat.},\,\rm{homog.}}={}&-\mathcal{L}_{\rm H}\big|_{\rm{stat.},\,\rm{homog.}} \\
        ={}& -\frac{c_1}{2}{(\mu\text{\bf H}\times \hat{\bm{n}})}^2 \\
        ={}& \frac{c_1}{2} \mu^2 \text{H}^2 \left( \hat n_z^2 - 1 \right)  \,,
    \end{split}
\end{align}
where we take the magnetic field to be aligned along the $z$-axis, $\text{\bf H} = \text{H} \:\hat{\bm z}$, a configuration we will assume in the rest of the work for simplicity. (A more general case is reported in Appendix~\ref{app:generalB}.) The Hamiltonian above is clearly minimized by any configuration such that $\hat n_z = 0$. Therefore, while in absence of magnetic field the background value of the order parameter can be aligned along any direction, now this freedom is restricted to the plane perpendicular to the magnetic field itself. We will then take our background value to be, for example, $\langle \hat{\bm n} \rangle = \hat{\bm y}$.

Following the standard coset construction ideas~\cite{callan1969structure,coleman1969structure}, the magnon fields can be parametrized as local broken transformations around equilibrium. Specifically, the background spontaneously breaks the rotations generated by $S_1$ and $S_3$, meaning that we can express the magnon fields as $\hat{\bm n}(x) = e^{i ( \theta^1(x) S_1 + \theta^3(x) S_3 )} \hat{\bm y}$. The corresponding quadratic Lagrangian is then,
\begin{align}
    \begin{split}
        \mathcal{L}_{\rm H}={}& \frac{c_1}{2} \bigg[ {\big(\dot\theta^1\big)}^2 -v_\theta^2 {\big(\bm \nabla \theta^1\big)}^2 - \mu^2 {\rm H}^2{\big(\theta^1\big)}^2 \\
        & + {\big(\dot\theta^3\big)}^2 - v_\theta^2 {\big(\bm\nabla\theta^3\big)}^2  + \mathcal{O}\big(\theta^4\big) \bigg] \,,
    \end{split}
\end{align}
and, the dispersion relations can be read off to be,
\begin{subequations}
\begin{align} \label{eq:isotropic}
    \omega_{q,-} ={}& v_\theta q \,,\\
    \omega_{q,+} ={}& \sqrt{\mu^2 {\rm H}^2 + v_\theta^2 q^2} \,.
\end{align}
\end{subequations}
A comment about this spectrum is in order.
In absence of magnetic field, the two modes are degenerate, and both gapless. This is because the background Hamiltonian of the system would be completely insensitive to the direction taken by $\langle \hat{\bm n} \rangle$. The two gapless modes are the Goldstone bosons reflecting this fact.

What happens at finite magnetic field? First of all, the existence of the gapless mode is the hallmark of the fact that the background Hamiltonian in Eq.~\eqref{eq:scHB} is still degenerate, but now only on the plane perpendicular to the $z$-axis. The gapless mode is the corresponding Goldstone boson. Secondly, the second mode, has a gap that is {\it universal}, meaning that it does not depend on the microscopic details of the system, which here are encoded in the effective coefficient $c_1$: it is solely dictated by the magnetic field. This is, instead, the hallmark of the so-called gapped Goldstones~\cite{Nicolis:2012vf,Nicolis:2013sga,Watanabe:2013uya,Cuomo:2020gyl}, which arise when a system has a finite density for one of its non-Abelian broken generators. This is indeed the case here for the broken charge $S_3$. In fact, the background value of the spin density, obtained from the SO(3) Noether current derived from the Lagrangian~\eqref{eq:Lagrangian for B field}, is given by
\begin{align} \label{eq:s}
    \langle \bm s \rangle = \big\langle c_1 \left[ \partial_t \hat{\bm n} \times \hat{\bm n} + \text{\bf H} - (\text{\bf H} \cdot \hat{\bm n}) \, \hat{\bm n} \right] \big\rangle = c_1 {\rm H} \hat{\bm z} \,.
\end{align}
As another check of this, one can verify that the Lagrangian~\eqref{eq:Lagrangian for B field} can be obtained from the Lagrangian~\eqref{eq:zero Lagrangian}, valid in absence of magnetic field, by replacing $\hat{\bm n} \to e^{- i \mu {\rm H} t S_3} \cdot \hat{\bm n}$, which shows that the background under consideration also breaks time translations, but preserves a combination of them with a spin rotation generated by $S_3$. In other words, the symmetry breaking pattern in the presence of an external magnetic field is ${\rm SO}(3) \times H \to \bar H = H + \mu \text{\bf H} \cdot \bm S$.\footnote{To keep the notation light, we omit the discrete symmetries.} One recognizes $\bar H$ to be precisely the modified Hamiltonian used to defined the ground state of a system at finite density for the operator $\text{\bf H}\cdot \bm S$. Crucially, gapped Goldstones are still {\it exact} Goldstone, meaning that they are needed to realize non-linearly the full SO(3) symmetry of the system.

%%%%%%%%%%%%%%%%%%%%%%%%%%%%%%%%%%%%%%

\subsection{EFT with anisotropies}\label{sec: EFT wit anisotropies}

\noindent As anticipated, we are interested in extending the EFT presented above to include the effects of tiny explicit breaking of the internal SO(3) symmetry, which are at the origin of the observed magnon gaps~\cite[e.g.,][]{hutchings_inelastic_1971,HutchingsSamuelsen}. As mentioned in the Introduction, we are interested in the case of ``easy plane'' antiferromagnets, which involve two distinct anisotropies. (For a simpler illustrative case, see Ref.~\cite{Brauner:2024juy}.)
In particular, the microscopic Hamiltonian is modified to~\cite{PhysRev.85.329,HutchingsSamuelsen}
\begin{align} \label{eq:Htot}
    H = H_{\rm H} + \sum_{\bm i} D_x \left( \bm{\mathcal{S}}_{\bm i}^x \right)^2 -  \sum_{\bm i} D_z \left( \bm{\mathcal{S}}_{\bm i}^z \right)^2 \,,
\end{align}
where $D_x \,, D_z>0$ are new anisotropic energies. The anisotropies along the $x$- and $z$-axes typically go under the name of, respectively, ``hard axis'' and ``easy axis'' anisotropy~\cite{rezende_introduction_2019}.

From the low energy viewpoint we again look at the symmetry breaking pattern. It is evident that the two anisotropies, if taken separately, would explicitly break the SO(3) spin rotations down to rotations around $\hat{\bm x}$, for the hard axis anisotropy, and around $\hat{\bm z}$, for the easy axis one. Since the additional terms in the Hamiltonian~\eqref{eq:Htot} are quadratic in the spins, they are also invariant under flipping of all the spins at the same time, $\bm{\mathcal{S}}_{\bm i} \to -\bm{\mathcal{S}}_{\bm i}$, which in turn corresponds to $\hat{\bm n} \to -\hat{\bm n}$. It follows, as it is probably reasonable to expect, that the additional terms in the effective Lagrangian are quadratic in the order parameter, i.e., 
\begin{align} \label{eq:n Lagrangian}
    \begin{split}
        \mathcal{L} ={}& \frac{c_1}{2}\Big[{(\partial_t\hat{\bm{n}}+\mu\text{\bf H}\times\hat{\bm{n}})}^2 -{v_\theta^2}{(\nabla_i\hat{\bm{n}})}^2 \\
        & + 2{\lambda_z}\hat{n}_z^2-2{\lambda_x}\hat{n}_x^2 \Big] \,,
    \end{split}
    \end{align}
where $\lambda_x\,,\lambda_z>0$ are two new effective coefficients, which can be determined in terms of the microscopic anisotropic energies (see Sec.~\ref{sec:matching}). Their sign has been chosen for later convenience. In particular, the last two terms indeed explicitly break the full SO(3) group.

\begin{figure}
\tikzset{every picture/.style={line width=0.75pt}} %set default line width to 0.75pt        

\begin{tikzpicture}[x=0.55pt,y=0.55pt,yscale=-1,xscale=1]
%uncomment if require: \path (0,473); %set diagram left start at 0, and has height of 473

%Straight Lines [id:da8746691297631699] 
\draw    (119,149) -- (227,149) ;
\draw [shift={(230,149)}, rotate = 180] [fill={rgb, 255:red, 0; green, 0; blue, 0 }  ][line width=0.08]  [draw opacity=0] (8.93,-4.29) -- (0,0) -- (8.93,4.29) -- cycle    ;
%Straight Lines [id:da2694826939649424] 
\draw    (119,149) -- (74.49,192.25) ;
\draw [shift={(72.33,194.34)}, rotate = 315.83] [fill={rgb, 255:red, 0; green, 0; blue, 0 }  ][line width=0.08]  [draw opacity=0] (8.93,-4.29) -- (0,0) -- (8.93,4.29) -- cycle    ;
%Straight Lines [id:da541907455255249] 
\draw    (119,149) -- (119,41) ;
\draw [shift={(119,38)}, rotate = 90] [fill={rgb, 255:red, 0; green, 0; blue, 0 }  ][line width=0.08]  [draw opacity=0] (8.93,-4.29) -- (0,0) -- (8.93,4.29) -- cycle    ;
%Shape: Circle [id:dp6851896455580211] 
\draw   (9.06,149) .. controls (9.06,88.28) and (58.28,39.06) .. (119,39.06) .. controls (179.72,39.06) and (228.94,88.28) .. (228.94,149) .. controls (228.94,209.72) and (179.72,258.94) .. (119,258.94) .. controls (58.28,258.94) and (9.06,209.72) .. (9.06,149) -- cycle ;
%Shape: Arc [id:dp10786627442614183] 
\draw  [draw opacity=0] (228.92,147.94) .. controls (228.93,148.3) and (228.94,148.65) .. (228.94,149) .. controls (228.94,175.79) and (179.72,197.5) .. (119,197.5) .. controls (58.8,197.5) and (9.91,176.16) .. (9.07,149.69) -- (119,149) -- cycle ; \draw   (228.92,147.94) .. controls (228.93,148.3) and (228.94,148.65) .. (228.94,149) .. controls (228.94,175.79) and (179.72,197.5) .. (119,197.5) .. controls (58.8,197.5) and (9.91,176.16) .. (9.07,149.69) ;  
%Shape: Arc [id:dp27292932986737073] 
\draw  [draw opacity=0][dash pattern={on 4.5pt off 4.5pt}] (9.07,148.37) .. controls (9.84,121.87) and (58.76,100.5) .. (119,100.5) .. controls (177.21,100.5) and (224.85,120.45) .. (228.69,145.71) -- (119,149) -- cycle ; \draw  [color={rgb, 255:red, 0; green, 0; blue, 0 }  ,draw opacity=0.44 ][dash pattern={on 4.5pt off 4.5pt}] (9.07,148.37) .. controls (9.84,121.87) and (58.76,100.5) .. (119,100.5) .. controls (177.21,100.5) and (224.85,120.45) .. (228.69,145.71) ;  
%Straight Lines [id:da5050386386458336] 
\draw [color={rgb, 255:red, 0; green, 116; blue, 255 }  ,draw opacity=1 ][line width=1.5]    (119,149) -- (119,42) ;
\draw [shift={(119,38)}, rotate = 90] [fill={rgb, 255:red, 0; green, 116; blue, 255 }  ,fill opacity=1 ][line width=0.08]  [draw opacity=0] (11.61,-5.58) -- (0,0) -- (11.61,5.58) -- cycle    ;
%Straight Lines [id:da09845822867498288] 
\draw [color={rgb, 255:red, 208; green, 2; blue, 27 }  ,draw opacity=1 ][line width=1.5]    (119,149) -- (226,149) ;
\draw [shift={(230,149)}, rotate = 180] [fill={rgb, 255:red, 208; green, 2; blue, 27 }  ,fill opacity=1 ][line width=0.08]  [draw opacity=0] (11.61,-5.58) -- (0,0) -- (11.61,5.58) -- cycle    ;
%Shape: Circle [id:dp6709109710770367] 
\draw  [fill={rgb, 255:red, 0; green, 0; blue, 0 }  ,fill opacity=1 ] (115.75,149) .. controls (115.75,147.21) and (117.21,145.75) .. (119,145.75) .. controls (120.79,145.75) and (122.25,147.21) .. (122.25,149) .. controls (122.25,150.79) and (120.79,152.25) .. (119,152.25) .. controls (117.21,152.25) and (115.75,150.79) .. (115.75,149) -- cycle ;
%Straight Lines [id:da7492429143254187] 
\draw [color={rgb, 255:red, 208; green, 2; blue, 27 }  ,draw opacity=1 ][line width=1.5]    (242,81.33) -- (242,17.34) ;
\draw [shift={(242,13.34)}, rotate = 90] [fill={rgb, 255:red, 208; green, 2; blue, 27 }  ,fill opacity=1 ][line width=0.08]  [draw opacity=0] (11.61,-5.58) -- (0,0) -- (11.61,5.58) -- cycle    ;

% Text Node
\draw (52.33,197.4) node [anchor=north west][inner sep=0.75pt]    {$\langle \hat{\bm{n}}_{x} \rangle $};
% Text Node
\draw (95,153) node [anchor=north west][inner sep=0.75pt]  [color={rgb, 255:red, 0; green, 117; blue, 255 }  ,opacity=1 ,rotate=-267.89]  {$ \langle \hat{\bm{n}} \rangle\,, {\rm H} \!<\! {\rm H}_{\rm{s.f.}}$};
% Text Node
\draw (252.67,43.73) node [anchor=north west][inner sep=0.75pt]  [color={rgb, 255:red, 208; green, 2; blue, 27 }  ,opacity=1 ]  {$\text{\bf H}$};
% Text Node
\draw (110,155) node [anchor=north west][inner sep=0.75pt]  [color={rgb, 255:red, 208; green, 2; blue, 27 }  ,opacity=1 ]  {$\langle \hat{\bm{n}} \rangle,\, {\rm H} \!>\! {\rm H}_{\rm{s.f.}}$};
% Text Node
\draw (235.33,138.4) node [anchor=north west][inner sep=0.75pt]    {$\langle \hat{\bm{n}}_{y} \rangle $};
% Text Node
\draw (102.33,6.4) node [anchor=north west][inner sep=0.75pt]    {$\langle \hat{\bm{n}}_{z} \rangle $};

\end{tikzpicture}
\caption{Background configuration in the simultaneous presence of an external magnetic field $\text{\bf H} = {\rm H} \hat{\bm{z}}$, and both easy and hard axis anisotropies. Depending on the magnitude of the external magnetic field ${\rm H}$, the system undergoes a phase transition, which, in the EFT framework, consists in a discontinuous change of the direction of the ground state.}
\label{fig:ground state}
\end{figure}
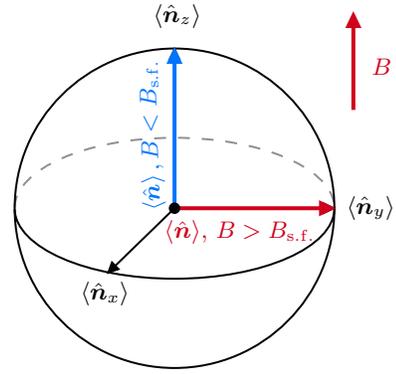

To define the ground state, we again consider static and homogeneous configurations in the Hamiltonian density,
\begin{align} \label{eq:Hforsf}
    \mathcal{H} \big|_{\rm{stat.},\,\rm{homog.}} ={}& -\frac{c_1}{2}\Big[(\mu\text{\bf H}\times \hat{\bm{n}})^2+2\lambda_z\hat{n}_z^2-2\lambda_x\hat{n}_x^2\Big] \notag \\
    ={}& \frac{c_1}{2} \Big[ (\mu^2 \text{H}^2 - 2 \lambda_z) \hat n_z^2 + 2\lambda_x \hat n_x^2 \Big] \,,
\end{align}
where we have dropped any irrelevant constant terms. Now, since $\lambda_x > 0$, the second term in the expression above is necessarily minimized by setting $\hat n_x = 0$. For the first term, instead, we have two possible configurations, depending on the relative strength between magnetic field effects and anisotropic ones. In particular, by defining a ``spin-flop'' field, $\text{H}_{\rm s.f.} \equiv \sqrt{2\lambda_z}/\mu$, we have
\begin{itemize}
    \item[$\circ$] for $ {\rm H}> {\rm H}_{\rm{s.f.}}$ the first terms is minimized by $\hat n_z = 0$, and the ground state is 
    \begin{align*}
        \langle \hat{\bm n} \rangle = \hat{\bm y} \,;
    \end{align*}
    In this phase, the background value of the spin density in Eq.~\eqref{eq:s} --- i.e., the total magnetization --- is $\langle \bm s \rangle = c_1 {\rm H} \hat{\bm z}$. This is what is typically called the ``spin-flop'' phase~\cite[e.g.,][]{rezende_introduction_2019}.
    
    \item[$\circ$] for ${\rm H}< {\rm H}_{\rm{s.f.}}$ the first term is minimized by $\hat n_z = 1$, and the ground state is 
    \begin{align*}
        \langle \hat{\bm n} \rangle = \hat{\bm z} \,.
    \end{align*}
    In this phase, instead, we have $\langle \bm s \rangle = 0$, which identifies this as the standard antiferromagnetic phase.
\end{itemize}
This situation is schematically represented in Fig.~\ref{fig:ground state}. 
Importantly, when the external magnetic field is varied across its spin-flop value, the order parameter, $\langle \hat{\bm n} \rangle$, is discontinuous. This is the hallmark of a first order phase transition~\cite{PhysRevB.95.104418,rezende_introduction_2019}.
Let us study these two phases separately, starting with the instance where the anisotropies are a small correction to what done so far, and then consider the case where they become dominant over the magnetic field. 

%%%%%%%%%%%%%%%%%%%%%%%%%%%%%%%%%%%%%%%%%%%

\subsubsection{Small anisotropies: the ${\rm H} > {\rm H}_{\rm s.f.}$ phase}

\noindent In this case the background is the same as in the absence of anisotropies, and we thus expect the explicit breaking effect to simply contribute to a small gap to the Goldstone modes. Indeed, by expanding again at quadratic order in the magnon fields, one gets,
\begin{align}
    \begin{split}
        \mathcal{L} ={}& \frac{c_1}{2} \bigg[ {( \dot \theta^1 )}^2 - v_\theta^2 {(\bm \nabla \theta^1)}^2 - (\mu^2 {\rm H}^2 - 2\lambda_z) {(\theta^1)}^2 \\
        & + {( \dot \theta^3 )}^2 - v_\theta^2 {(\bm \nabla \theta^3)}^2 - 2\lambda_x {(\theta^3)}^2 + \mathcal{O}\big(\theta^4\big) \bigg] \,.
    \end{split}
\end{align}
The corresponding gaps are given by,
\begin{subequations}\label{eq:Gaps for small anisotropies}
\begin{align}
    \omega_{q,+} ={}& \sqrt{2\lambda_x + v_\theta^2 q^2} \,, \\
    \omega_{q,-} ={}& \sqrt{\mu^2 {\rm H}^2 - 2\lambda_z + v_\theta^2 q^2} \,. \label{eq:omegaplussmall}
\end{align}
\end{subequations}
As anticipated, in this phase, the role of the anisotropies is that of turning the exact Goldstones described in Section~\ref{sec:isotropic} into {\it pseudo}-Goldstones. Moreover, the quadratic action above is simply a free theory for two independent real scalar fields. Consequently, it can be quantized in the standard way.

%%%%%%%%%%%%%%%%%%%%%%%%%%%%%%%%%%%%%%%%%%%

\subsubsection{Large anisotropies: the ${\rm H} < {\rm H}_{\rm s.f.}$ phase}

\noindent When the effects of the anisotropies are larger than those of the applied magnetic field, the background changes, and the system undergoes a phase transition. As an indication of this, the gap of the mode in Eq.~\eqref{eq:omegaplussmall} becomes imaginary, thus pointing to an instability. As anticipated, the new background is now $\langle \hat{\bm n} \rangle = \hat{\bm z}$. In this case, the broken generators are $S_1$ and $S_2$, and we must parametrize the magnon fields as $\hat{\bm{n}}(x) =e^{i\theta^a(x) S_a}\cdot\hat{\bm{z}}$, with $a=1,2$. Expanding Eq.~\eqref{eq:n Lagrangian} at quadratic order, one obtains
\begin{align} \label{eq: theta Lagrangian}
    \begin{split}
        \mathcal{L} ={}& \frac{c_1}{2} \bigg[ \big( \dot{\theta}^a - \mu {\rm H} \epsilon^{ab} \theta^b \big)^2 - v_\theta^2 \big( \bm \nabla \theta^a \big)^2 \\
        & - 2\lambda_z \big(\theta^a\big)^{2} - 2 \lambda_x \delta^{a2} \delta^{b2} \theta^a\theta^b + \mathcal{O}\big(\theta^4\big)  \bigg] \,.
    \end{split}
\end{align}
In this phase, the quadratic Lagrangian is not diagonal in the magnon fields, $\theta^1$ and $\theta^2$. To determine the spectrum, we must write the linerized equations of motion in Fourier space, i.e. $\mathcal{M}^{ab}(\omega,q) \theta^b(\omega,\bm q) = 0$, with
\begin{align} \label{eq:Quadratic Kernel}
    \begin{split}
        \mathcal{M}^{ab}(q,\omega)={}& c_1\bigg[\delta^{ab}\left(v_\theta^2{q}^2 -\omega^2+{2\lambda_z}-\mu^2 {\rm H}^2\right)\\
        &+2\lambda_x\delta^{a2}\,\delta^{b2}+2\,i\mu {\rm H}\,\omega\,\epsilon^{ab}\bigg]\,.
    \end{split}
\end{align}
The dispersion relations for the magnon modes are found by requiring that these equations of motion admit non-trivial solutions, that is to say, demanding that the determinant of the matrix above vanishes. After doing that, one obtains the following modes:
\begin{align} \label{eq:physical modes}
    \begin{split}
        \omega^2_{q,\alpha=\pm} ={}& \mu^2 {\rm H}^2+{\lambda_x+2\lambda_z} + v_\theta^2 q^2 \\
        & \pm\sqrt{ \lambda_x^2 + 4\mu^2 {\rm H}^2\left({\lambda_x+2\lambda_z} + v_\theta^2 q^2 \right)} \,.
    \end{split}
\end{align}
The corresponding gaps are
\begin{align} \label{eq:gaps}
    \begin{split}
        \omega^2_{0,\pm} ={}& \mu^2 {\rm H}^2+{\lambda_x+2\lambda_z} \\
        & \pm\sqrt{\lambda_x^2+4\mu^2 {\rm H}^2\left({\lambda_x+2\lambda_z}\right)}\,,
    \end{split}
\end{align}
which, for zero external field, are
\begin{subequations}
\begin{align}\label{eq:gap1}
    \omega_{0,+} \big|_{{\rm H}=0}  ={}& \sqrt{2(\lambda_x+\lambda_z)} \,,\\ \label{eq:gap2}
    \omega_{0,-}\big|_{{\rm H}=0} ={}& \sqrt{2\lambda_z} \,.
\end{align}
\end{subequations}

We can understand this spectrum using the following argument. For ${\rm H}=0$ and neglecting the easy-axis anisotropy (indeed, as we will see, for NiO crystals $\lambda_z\ll\lambda_x$), the ground state must have $\hat{n}_x=0$, i.e., performing rotations around the $x$-axis costs zero energy. There will thus be an exact Goldstone boson associated with this residual SO(2) invariance. However, introducing even
a small anisotropy along the $z$-axis breaks this residual invariance, forcing the ground state along the $z$-direction.\footnote{The high energy physicist would call this ``vacuum selection''.} We notice that the dispersion relations obtained in the isotropic case \eqref{eq:isotropic}
cannot be obtained taking the limit $\lambda_{x,z}\to0$ in \eqref{eq:physical modes}, since, in this limit, the background changes and the field parametrization must be re-evaluated. 

%%%%%%%%%%%%%%%%%%%%%%%%%%%%%%%%%%%%%%
%%%%%%%%%%%%%%%%%%%%%%%%%%%%%%%%%%%%%%

\section{Quantization of the EFT} \label{sec:quantization}

\noindent We are now ready to proceed with the quantization of our theory in the non-trivial case of ${\rm H} < {\rm H}_{\rm s.f.}$.
Lagrangian~\eqref{eq: theta Lagrangian} contains both a term with a single time derivative and one with two time derivatives. This makes it impossible to diagonalize the quadratic action by a local field redefinition. (See Ref.~\cite{Hui:2023pxc} for a recent discussion.) This can be seen, for example, from the fact that the eigenvectors of the matrix in Eq.~\eqref{eq:Quadratic Kernel} present non-analyticities in the frequency $\omega$. In order to deal with local fields, we then continue to work in the non-diagonal basis.
In this basis, the magnon fields, $\theta^a$, are not in one-to-one correspondence with the physical  magnon states, $|\bm q, \alpha\rangle$, which are (approximately) asymptotic states with dispersion relations $\omega_{q,\alpha}$, given in Eq.~\eqref{eq:physical modes}.\footnote{We remind the reader that the index $\alpha=\pm$ runs over the two possible magnon degrees of freedom.} This makes the quantization of the theory in the ${\rm H} < {\rm H}_{\rm s.f.}$ phase non-trivial.

In the following, we will denote with $a_{\bm q,\alpha}$ the annihilation operator associated with the single magnon state $\ket{\bm{q},\alpha}$, that is to say that,
\begin{align}
    a_{\bm q,\alpha}^\dagger\ket{0}=\ket{\alpha,\bm{q}} \,,
\end{align}
where the vacuum state is the one defined by the background configuration in presence of a magnetic field, as discussed in Sec.~\ref{sec: EFT wit anisotropies}.
We also adopt the relativistic normalization of the one-particle states, meaning the following commutation relation,
\begin{align}
    \left[a_{\bm q,\alpha},a^\dagger_{\bm p,\beta}\right]=2\omega_{\bm{q},\alpha}\,\delta_{\alpha\beta}(2\pi)^3\delta^{(3)}(\bm{q}-\bm{p})\,,
\end{align}
where the $\omega_{\bm{q},\alpha}$ are given in \eqref{eq:physical modes}. We will also work with canonically normalized fields, which are achieved by simply replacing $\theta^a \to \theta^a/\sqrt{c_1}$ in Eq.~\eqref{eq: theta Lagrangian}.

Since the quadratic Lagrangian is non-diagonal, we have that $\braket{0|\theta^a|\bm q,\alpha}\,\cancel{\propto}\,\delta^a_\alpha\,$. To proceed with the quantization procedure, we first write the mode expansion of our fields,
\begin{align}
    \begin{split}
        \theta^a(x) ={}& \sum_{\alpha=\pm}\int\frac{d^3q}{(2\pi)^32\omega_{q,\alpha}}e^{-i\omega_{q,\alpha}t+i\bm{q}\cdot\bm{x}}\,\mathcal{Z}^a_{q,\alpha}\,a_{\bm q,\alpha} \\
        & + \rm{h.c.} \,,
    \end{split}
\end{align}
where we introduced the overlap functions~\cite[e.g.,][]{Esposito:2020hwq,Hui:2023pxc,Cheung:2021yog,Creminelli:2023kze},  $\mathcal{Z}_{q,\alpha}^a$, which connect the $\theta^a$ fields to the physical magnon state, $\ket{\bm q,\alpha}$. They are defined as,
\begin{align}
    \braket{0|\theta^a(x)|\bm q,\alpha}=e^{-i\omega_{{q},\alpha}t+i\bm{q}\cdot\bm{x}} \, \mathcal{Z}^a_{q,\alpha}\,.
\end{align}
Since our theory is invariant under spatial rotations but not under boosts~\cite{Creminelli:2023kze}, the overlap functions will depend on $q$. To determine them, we impose the equal time commutators between the magnon fields and their conjugate momenta,
\begin{subequations}
\begin{align}
    \label{eq:commutator1}\left[\theta^a(\bm{x},t),\theta^b(\bm{y},t)\right]&=0\,,\\
    \label{eq:commutator2}\left[\theta^a(\bm{x},t),\pi^b(\bm{y},t)\right]&=i\,\delta^{ab}\,\delta^{(3)}(\bm{x}-\bm{y})\,,\\
    \label{eq:commutator3}\left[\pi^a(\bm{x},t),\pi^b(\bm{y},t)\right]&=0\,,
\end{align}
\end{subequations}
where $\pi^a \equiv \dot\theta^a-\mu \, {\rm H}\,\epsilon^{ab}\,\theta^b$. From the above commutators we get, respectively, the conditions,
\begin{subequations} \label{eq:condition}
\begin{align}
    \sum_{\alpha=\pm}\frac{\mathcal{Z}^a_{q,\alpha}\mathcal{Z}^{b \, *}_{q,\alpha}-\mathcal{Z}^{a *}_{q,\alpha}\mathcal{Z}^{b}_{q,\alpha}}{\omega_{q,\alpha}}&=0\,,\\
    \sum_{\alpha=\pm}\left[\mathcal{Z}^a_{q,\alpha}\mathcal{Z}^{b *}_{q,\alpha}+\mathcal{Z}^{a *}_{q,\alpha}\mathcal{Z}^{b}_{q,\alpha}\right]&=2\delta^{ab}\,,\\
    \sum_{\alpha=\pm}{\omega_{q,\alpha}}\left[\mathcal{Z}^a_{q,\alpha}\mathcal{Z}^{b *}_{q,\alpha}-\mathcal{Z}^{a *}_{q,\alpha}\mathcal{Z}^{b}_{q,\alpha}\right]&=4i\mu {\rm H}\epsilon^{ab}\,.
\end{align}
\end{subequations}
Imposing the equations of motion for ${\theta^a}$, we get the further constraint,
\begin{align}
    \mathcal{M}^{ab}(q,\omega_{q,\alpha})\mathcal{Z}^{b }_{q,\alpha}=0\,, \; \text{ for } \; \alpha=\pm \,,
\end{align}
where $\mathcal{M}^{ab}$ is defined in \eqref{eq:Quadratic Kernel}. 

We can use the equation above to solve for one component of the overlap functions in terms of the other. In particular, we get
\begin{align}\label{eq:Z1}
    \mathcal{Z}_{q,\pm}^2 ={}& i \left(\frac{v_\theta^2 q^2 - \mu^2 {\rm H}^2 + 2\lambda_z - \omega_{q,\pm}^2}{2\mu {\rm H} \:\omega_{q,\pm}} \right) \mathcal{Z}_{q,\pm}^1 \,.
\end{align}
Plugging this into the constraints reported in Eqs.~\eqref{eq:condition} one finally finds
\begin{align}\label{eq:Z2}
     {|\mathcal{Z}^1_{q,\pm}|} ={}& \sqrt{\frac{1}{2} \pm \frac{2\mu^2 {\rm H}^2 - \lambda_x}{2\sqrt{4\mu^2 {\rm H}^2 (v_\theta^2 q^2 + \lambda_x - 2\lambda_z) + \lambda_x^2}}} \,.
\end{align}
The overall phase, common to $\mathcal{Z}^1_{q,\alpha}$ and $\mathcal{Z}^2_{q,\alpha}$, is arbitrary.

We notice that, taking first the limit ${\rm H}\rightarrow0$, and then $\lambda_{z}\rightarrow0$, we get
\begin{align}
    \mathcal{Z}^a_{q,+}=\begin{pmatrix}0\\-i
    \end{pmatrix}\,,~~\mathcal{Z}^a_{q,-}=\begin{pmatrix}1\\0
    \end{pmatrix} \,,
\end{align}
as a check of the fact that, in this limit, the quadratic action reduces to a diagonal form.

\vspace{1em}

To make sure of the correctness of our results, in Appendix~\ref{app:polology}, we rederive the overlap functions using so-called ``polology'' arguments.

%%%%%%%%%%%%%%%%%%%%%%%%%%%%%%%%%%%%%%%%%%
%%%%%%%%%%%%%%%%%%%%%%%%%%%%%%%%%%%%%%%%%%

\section{Matching to a short distance theory} \label{sec:matching}

\begin{figure*}

\tikzset{every picture/.style={line width=0.75pt}} %set default line width to 0.75pt        

\begin{tikzpicture}[x=0.95pt,y=0.95pt,yscale=-1,xscale=1]
%uncomment if require: \path (0,331); %set diagram left start at 0, and has height of 331

%Shape: Square [id:dp24594762670753045] 
\draw   (37,54.5) -- (132,54.5) -- (132,149.5) -- (37,149.5) -- cycle ;
%Shape: Square [id:dp32809187344363044] 
\draw  [dash pattern={on 4.5pt off 4.5pt}] (79.1,21.5) -- (174.1,21.5) -- (174.1,116.5) -- (79.1,116.5) -- cycle ;
%Straight Lines [id:da534219772754658] 
\draw    (132,54.5) -- (174.1,21.5) ;
%Straight Lines [id:da8236902162065269] 
\draw    (37,54.5) -- (79.1,21.5) ;
%Straight Lines [id:da2695779445601718] 
\draw  [dash pattern={on 4.5pt off 4.5pt}]  (132,149.5) -- (174.1,116.5) ;
%Straight Lines [id:da6826450603353593] 
\draw  [dash pattern={on 4.5pt off 4.5pt}]  (37,149.5) -- (79.1,116.5) ;
%Straight Lines [id:da19462872288346822] 
\draw    (79.1,21.5) -- (174.1,21.5) ;
%Straight Lines [id:da22437118633171016] 
\draw    (174.1,25.57) -- (174.1,116.5) ;
%Straight Lines [id:da5522552499061] 
\draw    (174.1,116.5) -- (132,149.5) ;
%Straight Lines [id:da3538726071269429] 
\draw [color={rgb, 255:red, 74; green, 144; blue, 226 }  ,draw opacity=0.79 ][line width=1.5]  [dash pattern={on 5.63pt off 4.5pt}]  (79.35,21.64) -- (105.55,133) ;
%Straight Lines [id:da34731486564984704] 
\draw [color={rgb, 255:red, 208; green, 2; blue, 27 }  ,draw opacity=0.61 ][line width=1.5]  [dash pattern={on 5.63pt off 4.5pt}]  (103.83,38.15) -- (132,149.5) ;
%Shape: Square [id:dp22213886500823765] 
\draw   (216.27,35.86) -- (311.93,35.86) -- (311.93,131.53) -- (216.27,131.53) -- cycle ;
%Straight Lines [id:da2926867129612942] 
\draw [color={rgb, 255:red, 208; green, 2; blue, 27 }  ,draw opacity=1 ][line width=1.5]  [dash pattern={on 5.63pt off 4.5pt}]  (262.33,36.03) -- (311.93,131.53) ;
%Straight Lines [id:da19608731268971646] 
\draw [color={rgb, 255:red, 74; green, 144; blue, 226 }  ,draw opacity=1 ][line width=1.5]  [dash pattern={on 5.63pt off 4.5pt}]  (216.27,35.86) -- (265.86,131.36) ;
%Straight Lines [id:da8800439044352868] 
\draw    (244.83,71.19) -- (263.86,58) ;
\draw [shift={(265.5,56.86)}, rotate = 145.26] [fill={rgb, 255:red, 0; green, 0; blue, 0 }  ][line width=0.08]  [draw opacity=0] (9.6,-2.4) -- (0,0) -- (9.6,2.4) -- cycle    ;
%Straight Lines [id:da051589163490682655] 
\draw    (100.9,100.69) -- (114.67,96.13) ;
\draw [shift={(116.57,95.5)}, rotate = 161.67] [fill={rgb, 255:red, 0; green, 0; blue, 0 }  ][line width=0.08]  [draw opacity=0] (7.2,-1.8) -- (0,0) -- (7.2,1.8) -- cycle    ;
%Straight Lines [id:da7782893288459853] 
\draw    (100.9,100.69) -- (97.69,86.31) ;
\draw [shift={(97.25,84.36)}, rotate = 77.39] [fill={rgb, 255:red, 0; green, 0; blue, 0 }  ][line width=0.08]  [draw opacity=0] (7.2,-1.8) -- (0,0) -- (7.2,1.8) -- cycle    ;
%Straight Lines [id:da5064463834061517] 
\draw    (244.83,71.19) -- (232.76,50.59) ;
\draw [shift={(231.75,48.86)}, rotate = 59.64] [fill={rgb, 255:red, 0; green, 0; blue, 0 }  ][line width=0.08]  [draw opacity=0] (9.6,-2.4) -- (0,0) -- (9.6,2.4) -- cycle    ;
%Straight Lines [id:da20552703730921906] 
\draw [color={rgb, 255:red, 208; green, 2; blue, 27 }  ,draw opacity=1 ][line width=1.5]    (23.33,92.46) -- (16.21,63.68) ;
\draw [shift={(15.25,59.79)}, rotate = 76.1] [fill={rgb, 255:red, 208; green, 2; blue, 27 }  ,fill opacity=1 ][line width=0.08]  [draw opacity=0] (9.29,-4.46) -- (0,0) -- (9.29,4.46) -- cycle    ;
%Straight Lines [id:da04806619496905373] 
\draw [color={rgb, 255:red, 208; green, 2; blue, 27 }  ,draw opacity=1 ][line width=1.5]    (39.88,65.3) -- (35.15,47.56) ;
\draw [shift={(34.13,43.7)}, rotate = 75.1] [fill={rgb, 255:red, 208; green, 2; blue, 27 }  ,fill opacity=1 ][line width=0.08]  [draw opacity=0] (9.91,-4.76) -- (0,0) -- (9.91,4.76) -- (6.58,0) -- cycle    ;
%Straight Lines [id:da4145785285811876] 
\draw [color={rgb, 255:red, 208; green, 2; blue, 27 }  ,draw opacity=1 ][line width=1.5]    (155.93,96.3) -- (151.2,78.56) ;
\draw [shift={(150.18,74.7)}, rotate = 75.1] [fill={rgb, 255:red, 208; green, 2; blue, 27 }  ,fill opacity=1 ][line width=0.08]  [draw opacity=0] (9.91,-4.76) -- (0,0) -- (9.91,4.76) -- (6.58,0) -- cycle    ;
%Straight Lines [id:da8150805785150321] 
\draw [color={rgb, 255:red, 208; green, 2; blue, 27 }  ,draw opacity=1 ][line width=1.5]    (176.98,32.3) -- (172.25,14.56) ;
\draw [shift={(171.23,10.7)}, rotate = 75.1] [fill={rgb, 255:red, 208; green, 2; blue, 27 }  ,fill opacity=1 ][line width=0.08]  [draw opacity=0] (9.91,-4.76) -- (0,0) -- (9.91,4.76) -- (6.58,0) -- cycle    ;
%Straight Lines [id:da05525369568939342] 
\draw [color={rgb, 255:red, 74; green, 144; blue, 226 }  ,draw opacity=1 ][line width=1.5]    (81.2,28.58) -- (76.48,10.84) ;
\draw [shift={(82.23,32.45)}, rotate = 255.1] [fill={rgb, 255:red, 74; green, 144; blue, 226 }  ,fill opacity=1 ][line width=0.08]  [draw opacity=0] (9.91,-4.76) -- (0,0) -- (9.91,4.76) -- (6.58,0) -- cycle    ;
%Straight Lines [id:da1931346173238686] 
\draw [color={rgb, 255:red, 74; green, 144; blue, 226 }  ,draw opacity=1 ][line width=1.5]    (59.9,92.44) -- (55.18,74.7) ;
\draw [shift={(60.93,96.3)}, rotate = 255.1] [fill={rgb, 255:red, 74; green, 144; blue, 226 }  ,fill opacity=1 ][line width=0.08]  [draw opacity=0] (9.91,-4.76) -- (0,0) -- (9.91,4.76) -- (6.58,0) -- cycle    ;
%Straight Lines [id:da7860817507495683] 
\draw [color={rgb, 255:red, 74; green, 144; blue, 226 }  ,draw opacity=1 ][line width=1.5]    (38.85,152.37) -- (34.13,134.62) ;
\draw [shift={(39.88,156.23)}, rotate = 255.1] [fill={rgb, 255:red, 74; green, 144; blue, 226 }  ,fill opacity=1 ][line width=0.08]  [draw opacity=0] (9.91,-4.76) -- (0,0) -- (9.91,4.76) -- (6.58,0) -- cycle    ;
%Straight Lines [id:da7223401659733864] 
\draw [color={rgb, 255:red, 74; green, 144; blue, 226 }  ,draw opacity=1 ][line width=1.5]    (106.36,137) -- (102.33,119.53) ;
\draw [shift={(107.26,140.9)}, rotate = 257.02] [fill={rgb, 255:red, 74; green, 144; blue, 226 }  ,fill opacity=1 ][line width=0.08]  [draw opacity=0] (9.91,-4.76) -- (0,0) -- (9.91,4.76) -- (6.58,0) -- cycle    ;
%Straight Lines [id:da47158739091765245] 
\draw [color={rgb, 255:red, 74; green, 144; blue, 226 }  ,draw opacity=1 ][line width=1.5]    (175.95,123.44) -- (171.23,105.7) ;
\draw [shift={(176.98,127.3)}, rotate = 255.1] [fill={rgb, 255:red, 74; green, 144; blue, 226 }  ,fill opacity=1 ][line width=0.08]  [draw opacity=0] (9.91,-4.76) -- (0,0) -- (9.91,4.76) -- (6.58,0) -- cycle    ;
%Straight Lines [id:da47262350974084266] 
\draw [color={rgb, 255:red, 74; green, 144; blue, 226 }  ,draw opacity=1 ][line width=1.5]    (128.45,75.94) -- (123.72,58.2) ;
\draw [shift={(129.47,79.8)}, rotate = 255.1] [fill={rgb, 255:red, 74; green, 144; blue, 226 }  ,fill opacity=1 ][line width=0.08]  [draw opacity=0] (9.91,-4.76) -- (0,0) -- (9.91,4.76) -- (6.58,0) -- cycle    ;
%Straight Lines [id:da3274073500497967] 
\draw [color={rgb, 255:red, 208; green, 2; blue, 27 }  ,draw opacity=1 ][line width=1.5]    (106.37,48.25) -- (102,30.62) ;
\draw [shift={(101.04,26.73)}, rotate = 76.1] [fill={rgb, 255:red, 208; green, 2; blue, 27 }  ,fill opacity=1 ][line width=0.08]  [draw opacity=0] (9.91,-4.76) -- (0,0) -- (9.91,4.76) -- (6.58,0) -- cycle    ;
%Straight Lines [id:da8016574448045575] 
\draw [color={rgb, 255:red, 208; green, 2; blue, 27 }  ,draw opacity=1 ][line width=1.5]    (221.49,46.09) -- (212.86,29.2) ;
\draw [shift={(211.05,25.64)}, rotate = 62.94] [fill={rgb, 255:red, 208; green, 2; blue, 27 }  ,fill opacity=1 ][line width=0.08]  [draw opacity=0] (9.91,-4.76) -- (0,0) -- (9.91,4.76) -- (6.58,0) -- cycle    ;
%Straight Lines [id:da25364066102371563] 
\draw [color={rgb, 255:red, 208; green, 2; blue, 27 }  ,draw opacity=1 ][line width=1.5]    (269.93,94.54) -- (260.16,76.37) ;
\draw [shift={(258.27,72.85)}, rotate = 61.73] [fill={rgb, 255:red, 208; green, 2; blue, 27 }  ,fill opacity=1 ][line width=0.08]  [draw opacity=0] (9.91,-4.76) -- (0,0) -- (9.91,4.76) -- (6.58,0) -- cycle    ;
%Straight Lines [id:da1310393321379626] 
\draw [color={rgb, 255:red, 208; green, 2; blue, 27 }  ,draw opacity=1 ][line width=1.5]    (317.76,142.37) -- (307.99,124.2) ;
\draw [shift={(306.1,120.68)}, rotate = 61.73] [fill={rgb, 255:red, 208; green, 2; blue, 27 }  ,fill opacity=1 ][line width=0.08]  [draw opacity=0] (9.91,-4.76) -- (0,0) -- (9.91,4.76) -- (6.58,0) -- cycle    ;
%Straight Lines [id:da09859235430876678] 
\draw [color={rgb, 255:red, 208; green, 2; blue, 27 }  ,draw opacity=1 ][line width=1.5]    (267.86,46.53) -- (258.67,29.07) ;
\draw [shift={(256.81,25.53)}, rotate = 62.26] [fill={rgb, 255:red, 208; green, 2; blue, 27 }  ,fill opacity=1 ][line width=0.08]  [draw opacity=0] (9.91,-4.76) -- (0,0) -- (9.91,4.76) -- (6.58,0) -- cycle    ;
%Straight Lines [id:da46890538590207975] 
\draw [color={rgb, 255:red, 208; green, 2; blue, 27 }  ,draw opacity=1 ][line width=1.5]    (317.76,94.54) -- (307.99,76.37) ;
\draw [shift={(306.1,72.85)}, rotate = 61.73] [fill={rgb, 255:red, 208; green, 2; blue, 27 }  ,fill opacity=1 ][line width=0.08]  [draw opacity=0] (9.91,-4.76) -- (0,0) -- (9.91,4.76) -- (6.58,0) -- cycle    ;
%Straight Lines [id:da005817917297257003] 
\draw [color={rgb, 255:red, 74; green, 144; blue, 226 }  ,draw opacity=1 ][line width=1.5]    (220.21,138.85) -- (210.43,120.68) ;
\draw [shift={(222.1,142.37)}, rotate = 241.73] [fill={rgb, 255:red, 74; green, 144; blue, 226 }  ,fill opacity=1 ][line width=0.08]  [draw opacity=0] (9.91,-4.76) -- (0,0) -- (9.91,4.76) -- (6.58,0) -- cycle    ;
%Straight Lines [id:da5080088657263231] 
\draw [color={rgb, 255:red, 74; green, 144; blue, 226 }  ,draw opacity=1 ][line width=1.5]    (315.87,43.19) -- (306.1,25.02) ;
\draw [shift={(317.76,46.71)}, rotate = 241.73] [fill={rgb, 255:red, 74; green, 144; blue, 226 }  ,fill opacity=1 ][line width=0.08]  [draw opacity=0] (9.91,-4.76) -- (0,0) -- (9.91,4.76) -- (6.58,0) -- cycle    ;
%Straight Lines [id:da30127187772712527] 
\draw [color={rgb, 255:red, 74; green, 144; blue, 226 }  ,draw opacity=1 ][line width=1.5]    (268.82,136.98) -- (259.76,119.62) ;
\draw [shift={(270.67,140.53)}, rotate = 242.46] [fill={rgb, 255:red, 74; green, 144; blue, 226 }  ,fill opacity=1 ][line width=0.08]  [draw opacity=0] (9.91,-4.76) -- (0,0) -- (9.91,4.76) -- (6.58,0) -- cycle    ;
%Straight Lines [id:da2766048872428284] 
\draw [color={rgb, 255:red, 0; green, 0; blue, 0 }  ,draw opacity=0.48 ][line width=0.75]    (309.62,131.36) -- (222.1,131.36) ;
%Straight Lines [id:da5863613518687985] 
\draw [color={rgb, 255:red, 74; green, 144; blue, 226 }  ,draw opacity=1 ][line width=1.5]    (220.21,91.02) -- (210.43,72.85) ;
\draw [shift={(222.1,94.54)}, rotate = 241.73] [fill={rgb, 255:red, 74; green, 144; blue, 226 }  ,fill opacity=1 ][line width=0.08]  [draw opacity=0] (9.91,-4.76) -- (0,0) -- (9.91,4.76) -- (6.58,0) -- cycle    ;
%Straight Lines [id:da8667810008632943] 
\draw [color={rgb, 255:red, 0; green, 0; blue, 0 }  ,draw opacity=0.48 ][line width=0.75]    (216.27,38.48) -- (216.27,126) ;
%Straight Lines [id:da5123786174960476] 
\draw [color={rgb, 255:red, 0; green, 0; blue, 0 }  ,draw opacity=0.48 ][line width=0.75]    (311.91,48.91) -- (311.9,123.09) ;
%Straight Lines [id:da5339080872400979] 
\draw [color={rgb, 255:red, 0; green, 0; blue, 0 }  ,draw opacity=0.48 ][line width=0.75]    (285.55,35.82) -- (246.81,35.82) ;
%Straight Lines [id:da025530658609076218] 
\draw [color={rgb, 255:red, 208; green, 2; blue, 27 }  ,draw opacity=1 ][line width=1.5]    (208.07,94.01) -- (193.62,66.63) ;
\draw [shift={(191.75,63.09)}, rotate = 62.18] [fill={rgb, 255:red, 208; green, 2; blue, 27 }  ,fill opacity=1 ][line width=0.08]  [draw opacity=0] (9.29,-4.46) -- (0,0) -- (9.29,4.46) -- cycle    ;
%Straight Lines [id:da8749597625304217] 
\draw    (39,168.5) -- (130,168.5) ;
\draw [shift={(132,168.5)}, rotate = 180] [color={rgb, 255:red, 0; green, 0; blue, 0 }  ][line width=0.75]    (6.56,-1.97) .. controls (4.17,-0.84) and (1.99,-0.18) .. (0,0) .. controls (1.99,0.18) and (4.17,0.84) .. (6.56,1.97)   ;
\draw [shift={(37,168.5)}, rotate = 0] [color={rgb, 255:red, 0; green, 0; blue, 0 }  ][line width=0.75]    (6.56,-1.97) .. controls (4.17,-0.84) and (1.99,-0.18) .. (0,0) .. controls (1.99,0.18) and (4.17,0.84) .. (6.56,1.97)   ;
%Straight Lines [id:da5281631914966383] 
\draw    (337.93,129.53) -- (337.93,37.86) ;
\draw [shift={(337.93,35.86)}, rotate = 90] [color={rgb, 255:red, 0; green, 0; blue, 0 }  ][line width=0.75]    (6.56,-1.97) .. controls (4.17,-0.84) and (1.99,-0.18) .. (0,0) .. controls (1.99,0.18) and (4.17,0.84) .. (6.56,1.97)   ;
\draw [shift={(337.93,131.53)}, rotate = 270] [color={rgb, 255:red, 0; green, 0; blue, 0 }  ][line width=0.75]    (6.56,-1.97) .. controls (4.17,-0.84) and (1.99,-0.18) .. (0,0) .. controls (1.99,0.18) and (4.17,0.84) .. (6.56,1.97)   ;
%Shape: Triangle [id:dp34836359916671] 
\draw  [color={rgb, 255:red, 74; green, 144; blue, 226 }  ,draw opacity=0.72 ][fill={rgb, 255:red, 74; green, 144; blue, 226 }  ,fill opacity=0.19 ][dash pattern={on 4.5pt off 4.5pt}] (79.35,21.64) -- (172.28,116.46) -- (36.6,149.77) -- cycle ;
%Shape: Triangle [id:dp7727767743054512] 
\draw  [color={rgb, 255:red, 208; green, 2; blue, 27 }  ,draw opacity=0.52 ][fill={rgb, 255:red, 208; green, 2; blue, 27 }  ,fill opacity=0.12 ][dash pattern={on 4.5pt off 4.5pt}] (132.48,151.54) -- (36.3,55.35) -- (174.1,21.5) -- cycle ;
%Straight Lines [id:da4467142386245384] 
\draw [color={rgb, 255:red, 208; green, 2; blue, 27 }  ,draw opacity=1 ][line width=1.5]    (87.38,112.8) -- (82.65,95.06) ;
\draw [shift={(81.63,91.2)}, rotate = 75.1] [fill={rgb, 255:red, 208; green, 2; blue, 27 }  ,fill opacity=1 ][line width=0.08]  [draw opacity=0] (9.91,-4.76) -- (0,0) -- (9.91,4.76) -- (6.58,0) -- cycle    ;
%Straight Lines [id:da6930340770096048] 
\draw [color={rgb, 255:red, 208; green, 2; blue, 27 }  ,draw opacity=1 ][line width=1.5]    (134.88,160.3) -- (130.15,142.56) ;
\draw [shift={(129.13,138.7)}, rotate = 75.1] [fill={rgb, 255:red, 208; green, 2; blue, 27 }  ,fill opacity=1 ][line width=0.08]  [draw opacity=0] (9.91,-4.76) -- (0,0) -- (9.91,4.76) -- (6.58,0) -- cycle    ;
%Straight Lines [id:da2765551066581329] 
\draw [color={rgb, 255:red, 74; green, 144; blue, 226 }  ,draw opacity=1 ][line width=1.5]    (133.85,61.44) -- (129.13,43.7) ;
\draw [shift={(134.88,65.3)}, rotate = 255.1] [fill={rgb, 255:red, 74; green, 144; blue, 226 }  ,fill opacity=1 ][line width=0.08]  [draw opacity=0] (9.91,-4.76) -- (0,0) -- (9.91,4.76) -- (6.58,0) -- cycle    ;

% Text Node
\draw (7.07,80) node [anchor=north west][inner sep=0.75pt]  [color={rgb, 255:red, 208; green, 2; blue, 27 }  ,opacity=1 ]  {$\text{\bf H}$};
% Text Node
\draw (106.57,99.84) node [anchor=north west][inner sep=0.75pt]  [font=\normalsize]  {$\hat{\bm{x}}$};
% Text Node
\draw (97.71,70.88) node [anchor=north west][inner sep=0.75pt]  [font=\normalsize,color={rgb, 255:red, 0; green, 0; blue, 0 }  ,opacity=1 ]  {$\hat{\bm{z}}$};
% Text Node
\draw (261.93,61.16) node [anchor=north west][inner sep=0.75pt]  [font=\normalsize]  {$\hat{\bm{x}}$};
% Text Node
\draw (239.07,40.59) node [anchor=north west][inner sep=0.75pt]  [font=\normalsize,color={rgb, 255:red, 0; green, 0; blue, 0 }  ,opacity=1 ]  {$\hat{\bm{z}}$};
% Text Node
\draw (120,36) node [anchor=north west][inner sep=0.75pt]  [font=\normalsize,color={rgb, 255:red, 208; green, 2; blue, 27 }  ,opacity=1 ,rotate=-347.21]  {$\{111\}$};
% Text Node
% Text Node
\draw (20.,120.) node [anchor=north west][inner sep=0.75pt]  [font=\normalsize,color={rgb, 255:red, 0; green, 0; blue, 0 }  ,opacity=1,rotate=90 ]  {$[001]$};
\draw (40.,31.) node [anchor=north west][inner sep=0.75pt]  [font=\normalsize,color={rgb, 255:red, 0; green, 0; blue, 0 }  ,opacity=1,rotate=38 ]  {$[010]$};
\draw (110.,5.) node [anchor=north west][inner sep=0.75pt]  [font=\normalsize,color={rgb, 255:red, 0; green, 0; blue, 0 }  ,opacity=1,rotate=0 ]  {$[100]$};
% Text Node
% \draw (65.58,127.59) node [anchor=north west][inner sep=0.75pt]  [font=\normalsize,color={rgb, 255:red, 208; green, 2; blue, 27 }  ,opacity=1 ,rotate=-347.21]  {$\textcolor[rgb]{0.29,0.56,0.89}{\{111\}}$};
\draw (62,129) node [anchor=north west][inner sep=0.75pt]  [font=\normalsize,color={rgb, 255:red, 208; green, 2; blue, 27 }  ,opacity=1 ,rotate=-347.21]  {$\textcolor[rgb]{0.29,0.56,0.89}{\{111\}}$};
% Text Node
\draw (262,89.06) node [anchor=north west][inner sep=0.75pt]  [font=\normalsize,color={rgb, 255:red, 208; green, 2; blue, 27 }  ,opacity=1 ,rotate=-62.85]  {$\textcolor[rgb]{0.29,0.56,0.89}{[ 11}\textcolor[rgb]{0.29,0.56,0.89}{\overline{2}}\textcolor[rgb]{0.29,0.56,0.89}{]}$};
% Text Node
\draw (190.5,85) node [anchor=north west][inner sep=0.75pt]  [color={rgb, 255:red, 208; green, 2; blue, 27 }  ,opacity=1 ]  {$\text{\bf H}$};
% Text Node
\draw (283,39) node [anchor=north west][inner sep=0.75pt]  [font=\normalsize,color={rgb, 255:red, 208; green, 2; blue, 27 }  ,opacity=1 ,rotate=-62.85]  {$[{11\overline{2}}]$};
% Text Node
\draw (79.71,158.28) node [anchor=north west][inner sep=0.75pt]  [font=\normalsize,color={rgb, 255:red, 0; green, 0; blue, 0 }  ,opacity=1 ]  {$a$};
% Text Node
\draw (326.38,80.95) node [anchor=north west][inner sep=0.75pt]  [font=\normalsize,color={rgb, 255:red, 0; green, 0; blue, 0 }  ,opacity=1 ]  {$a$};

\end{tikzpicture}

\caption{The crystal structure and spin arrangements in NiO below the N\'eel temperature, $T_{\rm N}\simeq 523\text{ K}$, are characterized by a face-centered-cubic arrangements of $\rm{Ni}^{2+}$ ions. The $\rm{Ni}^{2+}$ ions align ferromagnetically along the $[11\overline 2 ]$ axis, indicated with bold dashed lines, lying within $\{111\}$ planes, depicted as colored triangular planes on the left. In this work, we consider the NiO sample in an external magnetic field, $\text{\bf H}$, that we choose oriented along the $[11\overline 2]$ axis. In the following we will adopt a reference frame where the $[ 111 ]$ and $[ 11\overline 2]$ axes, called hard and easy axes, coincide with the $x$- and $z$-axes, respectively.} \label{fig:crystal}
\end{figure*}
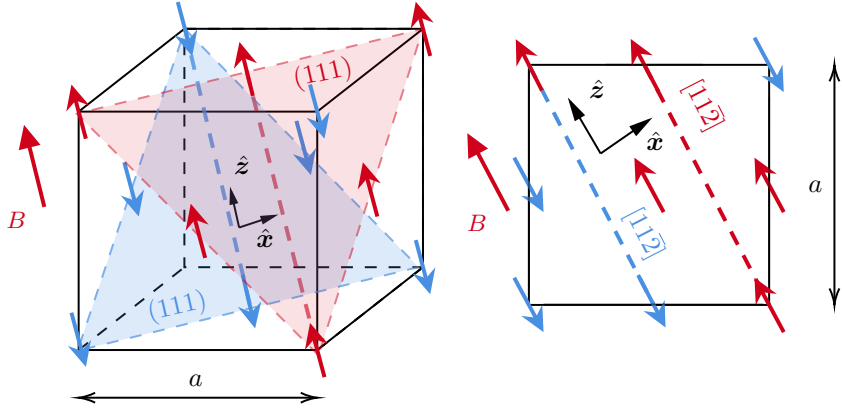

\noindent Let us now give explicit values of the effective coefficients appearing in the effective theory. To do that, we match some observable quantities obtained within the EFT with the same quantity as obtained from the short distance Hamiltonian. In particular, we will focus on the case of NiO, which is of particular phenomenological relevance~\cite[e.g.,][]{hutchings_inelastic_1971,milano2004effect,pheno}. 

In particular, the spin structure of NiO is dictated by that of its ${\rm Ni}^{2+}$ ions which, below its N\'eel temperature, $T_{\rm N} \simeq 523$~K, are ordered in stacked parallel planes. In each of these planes, the spins are aligned ferromagnetically, with each plane being antiferromagnetic with respect to the adjacent ones~\cite[e.g.,][]{rezende_introduction_2019}. In crystallographic jargon these are usually called $\{111\}$ planes --- see Fig.~\ref{fig:crystal}. Moreover, in each plane, the spins are aligned along the so-called $[ 11\bar 2]$ direction, as again showed in Figure~\ref{fig:crystal}. The spin structure just described is dictated by the presence of two anisotropies in spin space. The first one is along the direction perpendicular to the planes mentioned above, the so-called $[ 111 ]$ direction, and is responsible precisely for forcing the spins to lie on these planes. This is the hard axis anisotropy. The second one, instead, is in the $[ 11\bar 2]$ direction, and is responsible for the final orientation of the spins. This is the easy axis anisotropy. 

For this material, the spins are arranged in a face centered cubic lattice with the nearest, next-to-nearest, and next-to-next-to-nearest neighbors separated, respectively, by the following distances,
\begin{subequations}
\begin{align}
    |\bm i - \bm j| ={}& \frac{a}{\sqrt{2}} \quad \;\;\, \text{(nearest)} \,, \\
    |\bm i - \bm j| ={}& a \qquad \;\; 
    \, \text{(next-to-nearest)} \,, \\
    |\bm i - \bm j| ={}& \sqrt{\frac{3}{2}} a \quad \text{(next-to-next-to-nearest)} \,,
\end{align}
\end{subequations}
where $a$ is the cubic lattice parameter, see Fig.~\ref{fig:crystal}. As observed by neutron scattering experiments, the next-to-nearest spin-spin coupling is by far the dominant one~\cite{HutchingsSamuelsen,hutchings_inelastic_1971}, and the corresponding Heisenberg Hamiltonian is well approximated by
\begin{align}
    H ={}& 2 \sum_{\langle\bm i, \bm j\rangle} J_2 \, \bm{\mathcal{S}}_{\bm i} \cdot \bm{\mathcal{S}}_{\bm j}  + \sum_{\bm i} D_x \left( \bm{\mathcal{S}}_{\bm i}^x \right)^2 - \sum_{\bm i} D_z \left( \bm{\mathcal{S}}_{\bm i}^z \right)^2 \notag \\
    & + \mathcal{O}\left( J_1/J_2, J_3/J_2, \dots \right) \,,
\end{align}
where the sum over $\langle \bm i, \bm j \rangle$ runs only over lattice positions such that $|\bm i - \bm j| = a$, $J_2$ is the corresponding next-to-nearest-neighbor coupling, and $J_1$, $J_3$, and so on are the remaining sub-leading couplings.

Let us now proceed with the matching. The magnon speed of propagation, $v_\theta$, can be found from the dispersion relations in Eq.~\eqref{eq:physical modes} in the case of zero applied field, ${\rm H}=0$, since, for sufficiently large momenta, they become degenerate and linear, $\omega_{q,\alpha} \simeq v_\theta q$. The effective coefficient $c_1$, instead, is nothing but the so-called perpendicular magnetic susceptibility, $c_1 = \chi_\perp$, as one deduces from Eq.~\eqref{eq:s} (see also Refs.~\cite{Chakravarty:1987uk,manousakis1991spin}).
Its expression in terms of the short distance parameters, instead, has been found in Ref.~\cite{EspositoPavaskar} by computing the rate of emission of one magnon by neutron scattering both within the EFT, as well as within the short distance theory, as done in Ref.~\cite{lovesey}. The result is
\begin{align} \label{eq:c1}
    c_1 = \frac{4\mathcal{S}^2}{a v_\theta^2} J_2 \,.
\end{align}

\begin{figure}[t]
    \centering
    \includegraphics[width=0.95\columnwidth]{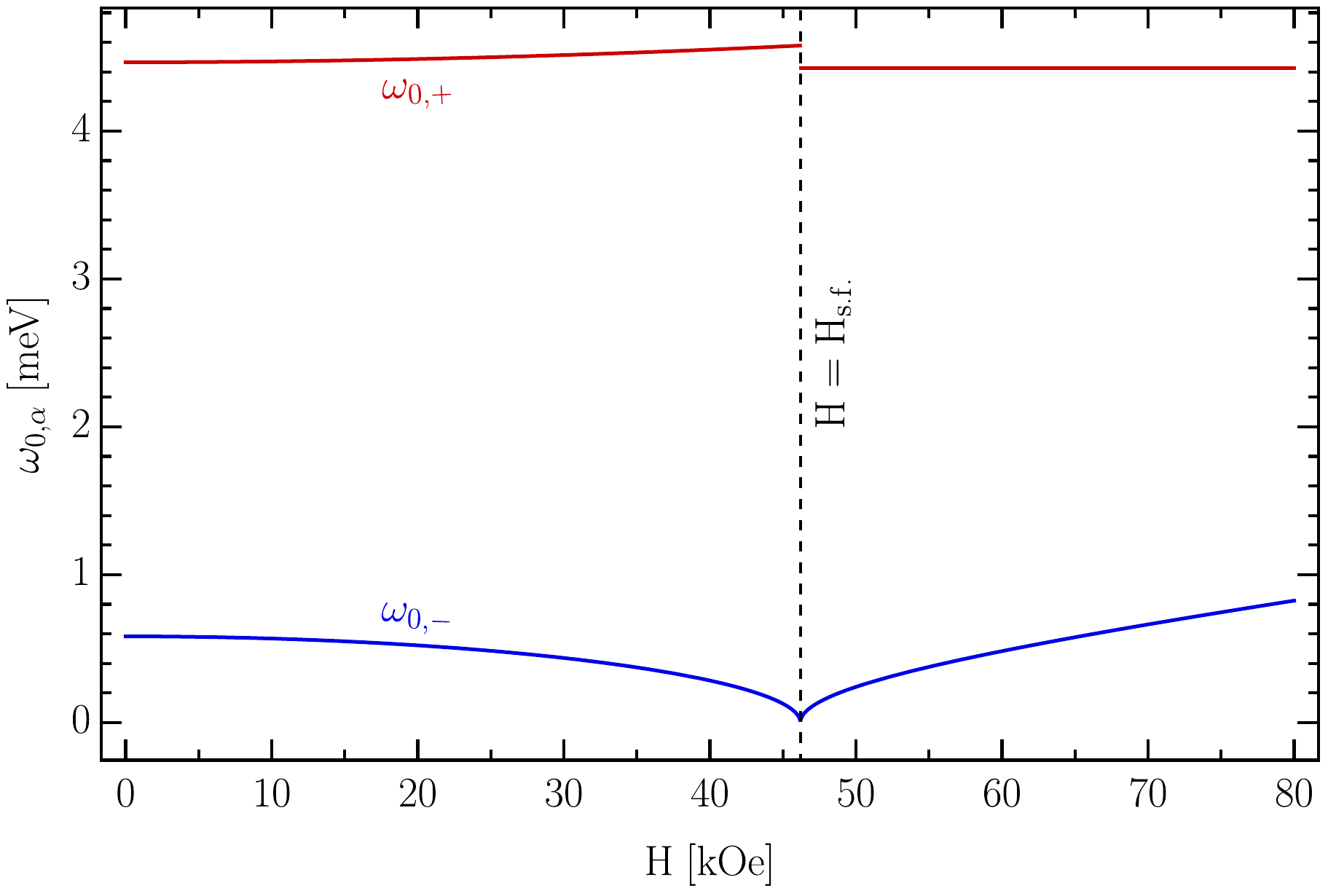}
    \caption{Gaps of the lowest lying magnon modes, $\omega_{0,\pm}$, as a function of the external magnetic field. Across the critical value, ${\rm H} = {\rm H}_{\rm s.f.}$, the lightest mode becomes gapless, while the higher one has a discontinuity.}
    \label{fig:gaps}
\end{figure}

The two anisotropic couplings, $
\lambda_x$ and $\lambda_z$, instead, can be determined by matching the values of the magnon gaps computed at zero magnetic field. In the short distance theory, these can be computed either by looking at the semiclassical solutions of the Landau-Lifshitz equations, or by actually quantizing the Hamiltonian and determining the spectrum of the corresponding magnon modes~\cite[e.g.,][]{rezende_introduction_2019}. Both methods return the same results:
\begin{subequations}
\begin{align} 
    \omega_{0,+}\big|_{{\rm H}=0} ={}& \mu \sqrt{2{\rm H}_E \left( {\rm H}_{Az} + {\rm H}_{Ax} \right)} \,, \\
    \omega_{0,-} \big|_{{\rm H}=0} ={}& \mu \sqrt{2 {\rm H}_E {\rm H}_{Az}} \,, 
\end{align}
\end{subequations}
where ${\rm H}_E$, ${\rm H}_{Az}$, and ${\rm H}_{Ax}$ are microscopic magnetic fields built out of the parameters of the Heisenberg Hamiltonian. Specifically, they are given by
\begin{align} \label{eq:Hmicro}
    {\rm H}_E = \frac{2 \mathcal{S} z J_2}{\mu} \,, \quad {\rm H}_{Ax} = \frac{2 \mathcal{S}D_x}{\mu} \,, \quad {\rm H}_{Az} = \frac{2\mathcal{S}D_z}{\mu} \,,
\end{align}
where $z$ is the number of next-to-nearest neighbors coupled by $J_2$, the so-called ``coordination number'', which for NiO is $z=6$. With this at hand, the low energy anisotropic couplings are found to be
\begin{subequations} \label{eq:lambdaxz}
\begin{align}
    \lambda_x ={}& \mu^2 {\rm H}_E {\rm H}_{Ax}\,,\\
    \lambda_z ={}& \mu^2 {\rm H}_E {\rm H}_{Az}\,.
\end{align}
\end{subequations}
In Table~\ref{tab:parameters}, we summarize the numerical values of both the short and long distance parameters for NiO. In Fig.~\ref{fig:gaps}, instead, we plot the gaps of the two lowest lying magnon modes of NiO as a function of the magnitude of the external magnetic field.

\begin{table*}[t]
    \centering
    \begin{tabular}{c | c c c c c c | c c c c}
         & \multicolumn{6}{c|}{\bf Short distance} & \multicolumn{4}{c}{\bf Long distance}  \\\hline\hline
         $\mu/\mu_{\rm B}$ & $\mathcal{S}$ & $a$ [\AA] & ${\rm H}_E$ [kOe] & ${\rm H}_{Ax}$ [kOe] & ${\rm H}_{Az}$ [kOe] & ${\rm H}_{\rm s.f.}$ [kOe] & $v_\theta$ & $c_1$ [MeV/\AA] & $\lambda_x$ [meV$^2$] & $\lambda_z$ [meV$^2$] \\\hline
         2.18 & 1 & 4.17 & 9684 & 6.35 & 0.11 & 46.3 & $1.3 \times 10^{-4}$ & 0.58 & 9.80 & 0.17 
    \end{tabular}
    \caption{Summary of the relevant parameters for NiO, both for the short distance Hamiltonian and for the long distance EFT. All short distance values are taken from Ref.~\cite{rezende_introduction_2019}, except for the lattice parameter, $a$, which is taken from Ref.~\cite{hutchings_inelastic_1971}. The propagation speeed, $v_\theta$, is fitted from the dispersion relations measured in Ref.~\cite{hutchings_inelastic_1971}. Finally, the effective coefficients $c_1$, $\lambda_x$ and $\lambda_z$ are determined from the matching conditions in Eqs.~\eqref{eq:c1} and \eqref{eq:lambdaxz}. In particular, in the expression for $c_1$ we traded the dominant coupling $J_2$ for the microscopic magnetic field ${\rm H}_E$. For the sake of the more high energy oriented reader, we recall that magnetic fields in a material are typically measured in oersted (Oe). In these units, the vacuum permeability is given by $\mu_0 = 10^{-4}$~T/Oe.}
    \label{tab:parameters}
\end{table*}

Note also that the value of the critical spin-flop magnetic field can also be determined in the short distance theory. In particular, it is given by~\cite{rezende_introduction_2019},
\begin{align}
    {\rm H}_{\rm s.f.} \simeq \sqrt{2 {\rm H}_E {\rm H}_{Az}} \,,
\end{align}
where the approximation comes from considering the limit ${\rm H}_{Ax,Az} \ll {\rm H}_E$. Using Eqs.~\eqref{eq:lambdaxz}, this reproduces exactly what was found independently within the EFT, as discussed below Eq.~\eqref{eq:Hforsf}.

%%%%%%%%%%%%%%%%%%%%%%%%%%%%%%%%%%%%%%
%%%%%%%%%%%%%%%%%%%%%%%%%%%%%%%%%%%%%%

\section{The regime of validity of the EFT}

\noindent As any effective theory, our EFT is valid only within certain conditions on the momenta and energies under consideration. Indeed, the Lagrangian in Eq.~\eqref{eq:n Lagrangian} is just the leading order in a systematic expansion in small momenta and energies. We then conclude our discussion by determining the limits of our EFT.

First of all, the theory is applicable as long as the wavelengths of interest are much larger than the typical distance characterizing the microscopic system. This translates into a cutoff on the momenta that one can consider: for momenta larger than this cutoff, one starts being sensitive to the short distance details of the material at hand. In a lattice, the natural momentum cutoff is given by the inverse lattice spacing, which means that our EFT is valid as long as the momenta are such that $q \ll 1/a$. As far as the energy cutoff is concerned, instead, the natural energy cutoff is given by the dominant spin-spin coupling, i.e., the theory is valid for $\omega \ll J_2$. Physically, this is the energy that one must provide to the system in order to be able to completely flip one of the spins.

Moreover, our theory also accounts for the effects of explicit symmetry breaking. This can be done consistently only as long as these effects are consistent with the regime of applicability of the EFT, and can therefore be included in the power counting. In detail, this can be quantified by requiring for the gaps induced by the anisotropies to be substantially smaller than the energy cutoff. This means to require that $\sqrt{\lambda_{x,z}} \ll J_2$. Using Eqs.~\eqref{eq:Hmicro} and \eqref{eq:lambdaxz}, this can be rephrased in terms of microscopic magnetic fields, corresponding to requiring ${\rm H}_{Ax,Az} \ll {\rm H}_E$. This condition is indeed well satisfied by our prototypical antiferromagnet, NiO (see Table~\ref{tab:parameters}).

Finally, since our setup also features an external magnetic field, we also need to make sure that this is sufficiently small not to cause drastic changes is the short distance structure of the material. Similarly as above, we can estimate the maximum magnetic field, beyond which the theory loses validity, as the value for which the gap of one of the magnon modes becomes comparable to the energy cutoff. When the magnetic field becomes large, it dominates the expression for the largest gap in Eq.~\eqref{eq:Gaps for small anisotropies}, which then become simply $\omega_{0,-} \simeq \mu {\rm H}$. 
When this is comparable to the energy cutoff, the theory is not applicable anymore. In other words, we must further make sure that the applied magnetic field satisfies ${\rm H} \ll J_2/\mu$. Note that, up to order one factors, this coincides with ${\rm H} \ll {\rm H}_E$.

%%%%%%%%%%%%%%%%%%%%%%%%%%%%%%%%%%%%%%
%%%%%%%%%%%%%%%%%%%%%%%%%%%%%%%%%%%%%%

\section{Outlook}

\noindent In this work, we have built a low energy EFT for anisotropic antiferromagnets. We showed that the introduction of small anisotropies induces the emergence of a number of interesting phenomena, especially in the presence of a magnetic field, which can nonetheless be studied analytically within the regime of applicability of the EFT.
This contributed to elucidate which of these phenomena are universal, meaning that they do not depend on the precise details of the short distance physics.

Moreover, an EFT description of this class of systems can help phrase the question in a language that can be understood from both a condensed matter and a high energy physics viewpoint. Such a description also makes it straightforward to compute magnon scattering amplitudes, which play an important role in determining the magnon lifetime in magnetic materials where traditional methods encounter difficulties~\cite{Dyson:1956zza}. This is also relevant in light of recent proposals to employ antiferromagnets as a target for light dark matter detection~\cite{EspositoPavaskar,pheno,Marsh:2018dlj}.

Various additional effects are, however, still left out of the present treatment. These are, for example, the inclusion of the long distance effects of dipolar interactions between spins. These are believed to be at the root of a richer spectrum observed in NiO, which features six additional light modes~\cite{milano2004effect}. These modes all have gaps substantially below the energy cutoff of our EFT and, thus, it should be possible to systematically account for them.
Furthermore, it is known that the magnon gaps can be varied by straining the antiferromagnetic sample~\cite{kim2022giant}. This requires a treatment which includes interactions with phonons in our EFT. We leave these and other interesting direction for future work.

%%%%%%%%%%%%%%%%%%%%%%%%%%%%%%%%%%%%%%%%%%
%%%%%%%%%%%%%%%%%%%%%%%%%%%%%%%%%%%%%%%%%%

\section{Acknowledgements}
\noindent We are grateful to Paolo~Barone, Matteo~Delladio, Bernard~van~Heck, Alberto~Nicolis, and Riccardo~Penco for useful discussions. We are also particularly thankful to Tom\'a\v{s}~Brauner
for extensive conversations on the role of magnetic fields in antiferromagnets, and to Jos\'e~Lorenzana for very insightful exchanges on the role of discrete symmetries. The work of SP is partially supported by DOE grant DE-SC0015655.

% \newpage

\appendix

\section{On the role of discrete symmetries in ferromagnets and antiferromagnets} \label{app:discrete}

\noindent We now take a moment to discuss systematically the role played by discrete symmetries and, in particular, how to employ them in the low energy EFT to discriminate between ferromagnets and antiferromagnets. In doing so, we will revisit some of the arguments that are typically found in the literature, rephrasing them in a way that, in our opinion, offers a viewpoint that is fully consistent with the proper low energy EFT ideology --- see also Ref.~\cite{Brauner:2024juy}.

Consider again a bipartite lattice of spins, as discussed in Sec.~\ref{sec:isotropic}. Regardless of what magnetic system one is actually considering, the presence of long range order is indicated by the existence of a given order parameter, say $\bm{\mathcal{O}}$, which transforms linearly under the ${\rm SO}(3)$ group of spin rotations, $\bm{\mathcal{O}} \to \mathcal{R} \cdot \bm{\mathcal{O}}$, and acquires a non-zero expectation value on the ground state, $\langle \bm{\mathcal{O}} \rangle \neq 0$. This fact alone tells us that the original spin rotation symmetry is spontaneously broken to the subgroup that leaves $\langle \bm{\mathcal{O}} \rangle$ unchanged. At this level, there is no distinction between ferromagnets and antiferromagnets. In both instances, their low energy EFT will be formulated in terms of a 3-vector field, $\hat{\bm n}(x)$, such that $|\hat{\bm n}(x)|=1$ and  $\langle \hat{\bm n}(x) \rangle \neq 0$. The difference arises when one wants to express the order parameter in terms of the degrees of freedom of the short distance theory, i.e., the spins of the material. In particular, one has
\begin{align*}
    \bm{\mathcal{O}} = \begin{cases} 
        \sum_{\bm i \in \mathds{A}} \bm{\mathcal{S}}_{\bm i} + \sum_{\bm i \in \mathds{B}} \bm{\mathcal{S}}_{\bm i} \equiv \bm{\mathcal{M}}\,, & \text{ferromagn.} \\
        \sum_{\bm i \in \mathds{A}} \bm{\mathcal{S}}_{\bm i} - \sum_{\bm i \in \mathds{B}} \bm{\mathcal{S}}_{\bm i} \equiv \bm{\mathcal{N}}\,, & \text{antiferromagn.}
    \end{cases} \,,
\end{align*}
where $\bm{\mathcal{M}}$ is the magnetization and $\bm{\mathcal{N}}$ is the N\'eel vector.

How about discrete symmetries? As mentioned again in Sec.~\ref{sec:isotropic}, both time reversal, $\mathcal{T}$, and the discrete rotation of $180^\circ$ in the plane containing the spins, $\mathcal{R}_\pi$, flip the direction of all the spins. But then, the order parameter also changes sign under either of these transformations,
\begin{align}
    \bm{\mathcal{O}} \xrightarrow[]{\mathcal{T}} - \bm{\mathcal{O}} \,, \qquad \bm{\mathcal{O}} \xrightarrow[]{\mathcal{R}_\pi} - \bm{\mathcal{O}} \,, \qquad \bm{\mathcal{O}} \xrightarrow[]{\mathcal{T} \, \mathcal{R}_\pi} \bm{\mathcal{O}} \,.
\end{align}
Again, this is true regardless of whether one is dealing with a ferromagnet or an antiferromagnet. It then follows that both these systems feature {\it the same symmetry breaking pattern}, i.e. ${\rm SO}(3) \times \mathcal{T} \to {\rm SO}(2) \times (\mathcal{T} \, \mathcal{R}_\pi)$.

The rules of the game to build a low energy EFT for either of these systems are then exactly the same. 
\begin{enumerate}
    \item Consider the field $\hat{\bm n}(x)$ such that,
        \begin{itemize} 
            \item[$\circ$] it rotates under ${\rm SO}(3)$, $\hat{\bm n}(x) \xrightarrow{{\rm SO}(3)} \mathcal{R} \cdot \hat{\bm n}(x)$;
            
            \item[$\circ$] it changes sign under either time reversal or the discrete rotation of $180^\circ$, $\hat{\bm n}(x) \xrightarrow{\mathcal{T}} - \hat{\bm n}(x)$ and $\hat{\bm n}(x) \xrightarrow{\mathcal{R}_\pi} - \hat{\bm n}(x)$;
            
            \item[$\circ$] it has unit norm and acquires a non-zero expectation value on the ground state, $\langle \hat{\bm n}(x) \rangle \neq 0$.
        \end{itemize}

        \item Write the most general theory for $\hat{\bm n}(x)$ that is invariant under the full ${\rm SO}(3)$ and under the combined action of $(\mathcal{T}\, \mathcal{R}_\pi)$.

        \item Organize the theory in a derivative expansion.
\end{enumerate}
Following this, the most general leading-order Lagrangian respecting these rules is the one reported also in~\cite{Pavaskar_2022},
\begin{align} \label{eq:Lgeneral}
    \mathcal{L} = \frac{c_1}{2} {(\partial_t \hat{\bm n})}^2 - \frac{c_2}{2} {(\nabla_i \hat{\bm n})}^2  + c_3 \, (\partial_t \phi) \cos \theta \,,
\end{align}
where $\theta(x)$ and $\phi(x)$ are the polar and azimuthal angles defining $\hat{\bm n} = (\sin\theta \cos\phi, \,\sin\theta\sin\phi,\,\cos\theta)$. As shown in Ref.~\cite{Pavaskar_2022}, under an infinitesimal ${\rm SO}(3)$ rotation, the Lagrangian above changes by a total derivative, thus ensuring the invariance of the theory (modulo possible non-trivial topologies).

What discriminates a ferromagnet from an antiferromagnet then? The difference between the two systems lies in the behavior of the {\it magnetization}. First of all, for a ferromagnet, the magnetization changes sign under time reversal or, equivalently, under $\mathcal{R}_\pi$. For an antiferromagnet, instead, it does not.
Moreover, the background value of the magnetization of a ferromagnet is non-zero, while it vanishes for an antiferromagnet.

Starting from the above low energy EFT, the magnetization is simply found as the temporal component of the Noether current associated to the original ${\rm SO}(3)$ symmetry, i.e., the spin density. From Eq.~\eqref{eq:Lgeneral}, one gets
\begin{align} \label{eq:sgeneral}
    \bm s = c_1 \, \partial_t \hat{\bm n} \times \hat{\bm n} + c_3 \, \hat{\bm n} \,.
\end{align}
Now it is clear that the only way that one has to reproduce the correct behavior of the magnetization described above must be if $c_1 = 0$ for ferromagnets, and $c_3 = 0$ for antiferromagnets, thus reducing the Lagrangian in Eq.~\eqref{eq:Lgeneral} to the known ones for these two systems.

\vspace{1em}

In light of the discussion above, we stress what we believe is an important conceptual point. The conditions $c_1 = 0$ or $c_3 = 0$ do not follow from different symmetry breaking patterns distinguishing ferromagnets from antiferromagnets. At low energies they have the same symmetry breaking pattern. What is happening here is different. The magnetization in Eq.~\eqref{eq:sgeneral}, which is a long distance observable, is an output of our EFT, not an input. By comparing the behavior of $\bm s$ under time reversal to that obtained from the microscopic theory of ferromagnets and antiferromagnets, we are rather performing a {\it matching} procedure which determines the values of the effective coefficients, as per usual. Alternatively, if we would have included the single-time derivative term in the Lagrangian for antiferromagnets, we would have arrived at the conclusion that they have a non-zero background magnetization, which is at odds with experiment.

%%%%%%%%%%%%%%%%%%%%%%%%%%%%%%%%%%%%%%%%%%

\section{Background and quadratic theory for a more general magnetic field} \label{app:generalB}

\noindent It is interesting to study what happens to the ground state of our theory, and to its quadratic Lagrangian, in the more general case of a magnetic field in the anisotropy plane, i.e., for $\text{\bf H} = {\rm H}\cos\theta \, \hat{\bm x} + {\rm H} \sin\theta \, \hat{\bm z}$. For simplicity we limit ourself to the range $\theta \in [0,\pi/2]$. In this instance, the Hamiltonian obtained from Eq.~\eqref{eq:n Lagrangian}, computed on static and homogeneous configurations, reads
\begin{align}
    \begin{split}
        \left.\mathcal{H}\right\vert_{{\rm{stat.,\,homog.}}}={}&\frac{c_1}{2}\big[(2\lambda_x+\mu^2{\rm H}^2\cos^2\theta)\hat{n}_x^2\\
        &\quad -(2\lambda_z-\mu^2{\rm H}^2\sin^2\theta)\hat{n}_z^2\\
        &\quad +2{\rm H}^2\hat{n}_x\hat{n}_z\sin\theta\cos\theta\big]\,,
    \end{split}
\end{align}
where again we omit irrelevant constant terms. One can verify that the background configuration that minimizes the Hamiltonian depends on the value taken by the projection of the magnetic field along the easy axis. In particular, for ${\rm H} \sin\theta > {\rm H}_{\rm s.f.}$ the background is $\langle \hat{\bm n} \rangle = \hat{\bm y}$, while for ${\rm H} \sin\theta < {\rm H}_{\rm s.f.}$ it is $\langle \hat{\bm n} \rangle = \hat{\bm z}$.

In this more general case, the quadratic Lagrangian is always non-diagonal, in each of the two phases. The Fourier space linear equations of motion for the magnon fields are again given by $\mathcal{M}^{ab}(\omega,q)\theta^b(\omega,\bm q)$, where now the kinetic matrix is given by
\begin{widetext}
\begin{subequations}
    \begin{align}
        \mathcal{M}^{ab} ={}& \begin{pmatrix}
            v_\theta^2q^2-\omega^2+\mu^2{\rm H}^2\sin^2\theta-2\lambda_z&-\mu ^2{\rm H}^2\cos\theta\sin\theta\\
            -\mu^2 {\rm H}^2\cos\theta \sin\theta&v_\theta^2q^2-\omega^2+\mu^2{\rm H}^2\cos^2\theta+2\lambda_x
        \end{pmatrix} \,, \qquad\qquad\quad\, \text{ for } {\rm H} \sin\theta > {\rm H}_{\rm s.f.} \,, \\
        \mathcal{M}^{ab} ={}& \begin{pmatrix}
            v_\theta^2q^2-\omega^2-\mu^2{\rm H}^2\sin^2\theta+2\lambda_z&2i\mu {\rm H}\omega\sin\theta\\
            -2i\mu {\rm H}\omega \sin\theta&v_\theta^2q^2-\omega^2+\mu^2{\rm H}^2\cos2\theta+2(\lambda_x+\lambda_z)
        \end{pmatrix} \,, \qquad \text{ for } {\rm H} \sin\theta < {\rm H}_{\rm s.f.} \,.
    \end{align}
\end{subequations}
\end{widetext}
Now, while neither of them is diagonal, the off-diagonal terms for the ${\rm H} \sin\theta > {\rm H}_{\rm s.f.}$ case do not depend on frequency. This means that the quadratic theory can be diagonalized by a simple, local field redefinition. In other words, one can simply work with a linear combination of the magnon fields, chosen to diagonalize the matrix.
For ${\rm H} \sin\theta < {\rm H}_{\rm s.f.}$, instead, the off-diagonal terms are frequency dependent, and one thus need to follow the same procedure discussed in Sec.~\ref{sec:quantization}.

The spectrum of the theory is found again by requiring that the matrices above have vanishing determinant. The final expressions can be found analytically, but they are considerably more involved than what was found in the main text. We thus omit them.

%%%%%%%%%%%%%%%%%%%%%%%%%%%%%%%%%%%%%%%%%%

\section{Overlap functions from “polology”} \label{app:polology}

\noindent To verify our procedure, we rederive the overlap functions with an alternative method, using the so-called “polology" technique~\cite{Weinberg:1995mt}, in a way analogous to what has been recently done in Ref.~\cite{Hui:2023pxc,Creminelli:2023kze}. 
As usual for such kind of manipulations, we will be using the completeness relation,
\begin{align}\label{eq:Completeness relation}
    \mathds{1}=\sum_{\alpha=\pm}\int\frac{d^3q}{(2\pi)^32\omega_{{q},\alpha}}\ket{\bm q,\alpha}\bra{\bm q,\alpha}+\dots \,,
\end{align}
where the ellipsis stands for multi-particle states. The latter will not play any role as long as one is interested in {\it tree level} $S$-matrix elements, for which one considers the on-shell limit of the external {\it one}-particle states. Inserting the above relation in the time-ordered two-point function (or, alternatively, the correlator) of the magnon field, one gets
\begin{widetext}
\begin{align}
    \begin{split}
        \braket{0|\mathcal{T}\left\{{\theta^a}(\bm{x},t){\theta^b}(\bm{y},t')\right\}|0} ={}& \braket{0|{\theta^a}(\bm{x},t){\theta^b}(\bm{y},t')|0}\theta(t-t') +\braket{0|{\theta^b}(\bm{y},t'){\theta^a}(\bm{x},t)|0}\theta(t'-t) \\
        ={}& \sum_{\alpha=\pm} \int  \frac{d^3q}{(2\pi)^3 2\omega_{q,\alpha}} \big[ \langle 0 | \theta^a(\bm x, t) |\bm q, \alpha \rangle \langle \bm q, \alpha | \theta^b(\bm y, t') | 0 \rangle \theta(t-t') \\
        & + \langle 0 | \theta^b(\bm y, t') |\bm q, \alpha \rangle \langle \bm q, \alpha | \theta^a(\bm x, t) | 0 \rangle \theta(t'-t)  \big] + \dots \\
        ={}& \int\frac{d^4 q}{(2\pi)^4}e^{-i \omega(t-t')+i\bm{q}\cdot(\bm{x}-\bm{y})} \sum_{\alpha=\pm}\frac{1}{2\omega_{q,\alpha}}\left[\frac{i\mathcal{Z}^a_{q,\alpha} \mathcal{Z}^{b*}_{q,\alpha}}{\omega-\omega_{q,\alpha}+i\varepsilon}-\frac{i\mathcal{Z}^{a*}_{q,\alpha} \mathcal{Z}^b_{q,\alpha}}{\omega+\omega_{q,\alpha}-i\varepsilon}\right] + \dots \,, \\
        \equiv{}& \int\frac{d^4 q}{(2\pi)^4}e^{-i \omega(t-t')+i\bm{q}\cdot(\bm{x}-\bm{y})} \mathcal{K}^{ab}(\omega, q) + \dots \,,
    \end{split}
\end{align}
\end{widetext}
where $d^4q=d\omega d^3q$, $\ket{0}$ is the vacuum in the presence of external magnetic field and anisotropies (see Sec.~\ref{sec: EFT wit anisotropies}), and we used the identity $\theta(t)= \int\frac{d\omega}{2\pi}\frac{i}{\omega+i\varepsilon}e^{-i\omega t}$. 

Now, at tree level, the Fourier transform of the two-point function is the inverse of the kinetic matrix, which we reported in Eq.~\eqref{eq:Quadratic Kernel}, i.e.,
\begin{align} \label{eq:Kab}
    \mathcal{K}^{ab}(\omega,{q})={}&-i\left[\mathcal{M}^{ab}(\omega,{q})\right]^{-1} \notag \\
    ={}& \frac{-i}{\det\left[\mathcal{M}(\omega,{q})\right]} \\
    & \times \Big(\delta^{ab}\big[ v_\theta^2 q^2 -\mu^2 {\rm H}^2 -\omega^2+{2(\lambda_z+\lambda_z)}\big] \notag \\ 
    & \quad -\delta^{a2}\,\delta^{b2}{2\lambda_x}-2i\mu {\rm H} \omega\,\epsilon^{ab}\Big)\,. \notag
\end{align}
Since this is an identity between Fourier coefficients, it must hold true for any value of $\omega$, and should thus be solved by comparing terms with the same powers of the frequency. Specifically, setting $a=b=1,2$, we can extract  the absolute values of the overlap functions. Indeed, for $a=b=1$, we get ${|\mathcal{Z}_{q,-}^1|}^2+{|\mathcal{Z}_{q,+}^1|}^2=1$ and
\begin{align}
    \sum_{\alpha=\pm}|\mathcal{Z}_{q,\alpha}^1|^2\omega_{q,-\alpha}^2 = v_\theta^2 q^2 - \mu^2 {\rm H}^2
    +{2(\lambda_z+\lambda_z)}\,,
\end{align}
where $\omega_{-\alpha}=\omega_{\mp}$, for $\alpha=\pm$.
Analogously, for $a=b=2$, we get $|\mathcal{Z}_{q,-}^2|^2+|\mathcal{Z}_{q,+}^2|^2=1$ and
\begin{align}
    \sum_{\alpha=\pm}|\mathcal{Z}_{q,\alpha}^2|^2\omega_{q,-\alpha}^2=v_\theta^2q^2-\mu^2 {\rm H}^2
    +{2\lambda_z}\,.
\end{align}
Therefore we find that
\begin{align*}
     {|\mathcal{Z}^1_{q,\pm}|} ={}& \sqrt{\frac{1}{2} \pm \frac{2\mu^2 {\rm H}^2 - \lambda_x}{2\sqrt{4\mu^2 {\rm H}^2 (v_\theta^2 q^2 + \lambda_x - 2\lambda_z) + \lambda_x^2}}} \,, \\
     {|\mathcal{Z}^2_{q,\pm}|} ={}& \sqrt{\frac{1}{2} \pm \frac{2\mu^2 {\rm H}^2 + \lambda_x}{2\sqrt{4\mu^2 {\rm H}^2 (v_\theta^2 q^2 + \lambda_x - 2\lambda_z) + \lambda_x^2}}} \,,
\end{align*}
in agreement with Eqs.~\eqref{eq:Z1} and \eqref{eq:Z2}.
To determine the relative phase between $\mathcal{Z}^1_{q,\alpha}$ and $\mathcal{Z}^2_{q,\alpha}$ we write them as
\begin{align}
    \mathcal{Z}_{q,\alpha}^a=\left|\mathcal{Z}_{q,\alpha}^a\right| e^{i\varphi_\alpha^a} \,.
\end{align}
By taking $a=1$, $b=2$ in Eq.~\eqref{eq:Kab}, we obtain the following equation for the relative phase, $\Delta \varphi_\alpha \equiv \varphi_\alpha^1 - \varphi_\alpha^2$,
\begin{align}\label{eq:Phases}
    \begin{split}
        &\sum_{\alpha}\frac{i|\mathcal{Z}_{q,\alpha}^1||\mathcal{Z}_{q,\alpha}^2|}{\omega_{q,\alpha}\left(\omega^2-\omega_{q,\alpha}^2\right)}\left[i\,\omega\sin\Delta\varphi_\alpha+\omega_{q,\alpha}\cos\Delta\varphi_\alpha\right]\\
        &\hspace{3cm}=-\frac{2\mu {\rm H} \omega}{\det\left[\mathcal{M}(\omega,\bm{q})\right]}\,.
    \end{split}
\end{align}
Equating again terms with the same powers of $\omega$, we find $\Delta\varphi_\pm=\pm\pi/2$, which again agrees with Eq.~\eqref{eq:Z1}.

%%%%%%%%%%%%%%%%%%%%%%%%%%%%%%%%%%%%%%%%%%
%%%%%%%%%%%%%%%%%%%%%%%%%%%%%%%%%%%%%%%%%%

% \bibliographystyle{apsrev4-1}
\bibliography{biblio.bib}

%%%%%%%%%%%%%%%%%%%%%%%%%%%%%%%%%%%%%%%%%%%%%%%%%%%

\end{document}